\documentclass[prc,preprint,showpacs,showkeys,nofootinbib]{revtex4}%
\usepackage{graphicx}
\usepackage{bm}
\usepackage{epsfig}
\usepackage{amsmath}
\usepackage{amsfonts}
\usepackage{amssymb}%
\begin{document}
\title{A computer program for nuclear scattering at intermediate and high energies}
\author{C.A. Bertulani, C.M. Campbell and T. Glasmacher }
\address{National Superconducting Cyclotron Laboratory, Michigan State University,
E. Lansing, MI 48824}
\date{\today}

\begin{abstract}
A computer program is presented which calculates the elastic and inelastic
scattering in intermediate and high energy nuclear collisions. A
coupled-channels method is used for Coulomb and nuclear excitations of E1, E2,
E3, M1, and M2 multipolarities, respectively. The program applies to an
arbitrary nucleus, specified by the spins and energies of the levels and by
reduced matrix elements. For given bombarding conditions, the angular
distribution of elastic and inelastic scattered particles and angular
distributions of gamma-rays from the excited nucleus are computed.

\end{abstract}
\maketitle

\draft

\narrowtext

\section{PROGRAM SUMMARY}

\begin{enumerate}
\item \textit{Title of program: }DWEIKO (Distorted Wave EIKOnal Approximation)

\textit{Program obtainable from:} CPC Program Library, Queen's University of
Belfast, N. Ireland

\textit{Computers:} The code has been created on an IBM-PC, but
also runs on UNIX machines.

\textit{Operating systems:} WINDOWS or UNIX

\textit{Program language used:} Fortran-77

\textit{Memory required to execute with typical data:} 8 Mbytes of RAM memory
and 1 MB of hard disk space

\textit{No. of bits in a word: }64

\textit{Memory required for test run with typical data:} 1 MB

\textit{No. of bytes in distributed program, including test data, etc.:} 82238

\textit{Distribution format:} ASCII

\textit{Keywords:} Elastic scattering; Coulomb excitation; Relativistic
collisions; Coupled-channels; Nuclear excitation

\textit{Nature of physical problem:} The program calculates elastic scattering
differential cross sections, probabilities, and cross sections for inelastic
scattering in nuclear collisions at intermediate and high energies
($E_{lab}\geq50$ MeV/nucleon). It is particularly useful in the analysis of
experiments with stable and unstable nuclear beams running at several
intermediate-energy heavy ion accelerators around the world.

\textit{Method of solution:} Eikonal wavefunctions are used for
the scattering. For each ``impact parameter" entering the
scattering matrix elements, one solves coupled-channels equations
for the time-dependent Coulomb + nuclear field expanded into
multipoles. A four-point Runge-Kutta procedure is used to solve
the coupled-channels equations. The elastic scattering is
calculated purely with the eikonal approximation. The
coupled-channels is a separate calculation for the inelastic
amplitudes. The inelastic couplings, therefore, have no effect on
the obtained elastic scattering cross sections.

\textit{Typical running time:} Almost all the CPU time is consumed by the
solution of the coupled-channels equations. It is about 2 min on a 1GHz Intel
P4-processor machine for the inclusion of 5 nuclear states.
\end{enumerate}

\section{LONG WRITE-UP}

\subsection{Introduction}

The eikonal approximation is very useful in the study of nucleus-nucleus
scattering at high energies \cite{Gl59}. In the Distorted Wave Born
Approximation (DWBA) the transition amplitude for the reaction $A(a,b)B$
involves a matrix element of the form \cite{BHM02}
\begin{equation}
T_{DWBA}=\int\Psi^{(-)*}_{\beta}(\mathbf{r})\langle b,B |U_{int}(\mathbf{r}%
)|a,A \rangle\Psi^{(+)}_{\alpha}(\mathbf{r}) d^{3}r_{\alpha}d^{3}r_{\beta
},\label{intro1}%
\end{equation}
where $U_{int}(\mathbf{r})$ is the interaction potential, and $\Psi_{\alpha}$
($\Psi_{\beta}$) is the scattering wave function in the entrance (exit)
channel, $\alpha=a+A$ ($\beta=b+B$). $\langle a,A|$ and $\langle b,B|$ are the
initial and final intrinsic wavefunctions of the system, respectively. Using
the eikonal approximation for the wave functions one has \cite{Gl59}
\begin{equation}
\Psi^{(-)\ast}(\mathbf{r})\Psi^{(+)}(\mathbf{r})\simeq\exp\left\{
i\mathbf{q.r}+i\chi(\mathbf{b})\right\}  \ , \label{intro2}%
\end{equation}
where $\chi(\mathbf{b})$ is the eikonal phase, given by
\begin{equation}
\chi(\mathbf{b})=-{\frac{1}{\hbar\mathrm{v}}}\int_{-\infty}^{\infty
}dz\ U_{opt}(\mathbf{r})\ .\label{intro3}%
\end{equation}
The conditions of validity of the eikonal approximation are: (a) forward
scattering, i.e., $\theta\ll1$ radian, and (b) small energy transfers from the
bombarding energy to the internal degrees of freedom of the projectile, or
target. Both conditions apply perfectly well to direct processes in nuclear
scattering at $E_{lab}\geq50$ MeV/nucleon \cite{Gl59}.

In the above equation, $U_{opt}(\mathbf{r})$ is the optical potential, with
$r=\sqrt{b^{2}+z^{2}}$, where $b$ can be interpreted as the impact parameter.
For the Coulomb part of the optical potential this integral diverges. One
solves this by using $\chi=\chi_{N}+\chi_{C}$, where $\chi_{N}$ is given by
the equation above without the Coulomb potential and writing the Coulomb
eikonal phase, $\chi_{C}$ as
\begin{equation}
\chi_{C}(b)=2\eta\ln(kb)\ ,\label{intro4}%
\end{equation}
where $\eta=Z_{1}Z_{2}e^{2}/\hbar\mathrm{v}$, $Z_{1}$ and $Z_{2}$ are the
charges of projectile and target, respectively, $v$ is their relative
velocity, $k$ their wavenumber in the center of mass system. Eq. \ref{intro4}
reproduces the exact Coulomb scattering amplitude when used in the calculation
of the elastic scattering with the eikonal approximation \cite{BHM02}:
\begin{equation}
f_{C}(\theta)={\frac{Z_{1}Z_{2}e^{2}}{2\mu v^{2}\ \sin^{2}(\theta/2)}}%
\ \exp\Big\{-i\eta\ \ln\Big[\sin^{2}(\theta/2)\Big]+i\pi+2i\phi_{0}%
\Big\}\label{fctheta}%
\end{equation}
where $\phi_{0}=arg\Gamma(1+i\eta/2)$. This is convenient for the numerical
calculations since, as shown below, the elastic scattering amplitude can be
written with the separated contribution of the Coulomb scattering amplitude.
Then, the remaining integral (the second term on the r.h.s. of eq.
\ref{simplif}) converges rapidly for the scattering at forward angles.

Although the Coulomb phase in eq. \ref{intro4} diverges at $b=0$, this does
not pose a real problem, since the strong absorption suppresses the scattering
at small impact parameters. One can correct it for a finite charge
distribution of the nucleus \cite{Fa72}. For example, assuming a uniform
charge distribution with radius $R$ the Coulomb phase becomes
\begin{align}
\chi_{C}(b)  & =2\eta\ \bigg\{\Theta(b-R)\ \ln(kb)+\Theta(R-b)\Big[\ln
(kR)+\ln(1+\sqrt{1-b^{2}/R^{2}})\nonumber\\
& -\sqrt{1-b^{2}/R^{2}}-{\frac{1}{3}}(1-b^{2}/R^{2})^{3/2}%
\Big]\bigg\}\ ,\label{chico_0}%
\end{align}
where $\Theta$ is the step function. This expression is finite for $b=0$,
contrary to eq. \ref{intro4}. If one assumes a gaussian distribution of charge
with radius $R$, appropriate for light nuclei, the Coulomb phase becomes
\begin{equation}
\chi_{C}(b)=2\eta\ \Big\{\ln(kb)+{\frac{1}{2}}E_{1}(b^{2}/R^{2}%
)\Big\}\ ,\label{chico2}%
\end{equation}
where the error function $E_{1}$ is defined as
\begin{equation}
E_{1}(x)=\int_{x}^{\infty}{\frac{e^{-t}}{t}}\ dt\ .\label{chico3}%
\end{equation}
This phase also converges, as $b\rightarrow0$. The cost of using
the expressions \ref{chico_0} and \ref{chico2} is that the Coulomb
scattering amplitude becomes more complicated than \ref{fctheta}.
Moreover, we have verified that the elastic and inelastic
scattering cross sections change very little by using eqs.
\ref{chico_0} or \ref{chico2}, instead of eq. \ref{intro4}.

The simplification introduced by the eikonal approximation is huge, as one
avoids the calculation of scattering wavefunctions by solving numerically the
Schr\"odinger equation for each partial wave, as is done DWBA for nuclear
scattering at low energies. For computer programs appropriate for nuclear
scattering at low energies, see, e.g., the codes FRESCO \cite{Ian1} and DWUCK4
\cite{KR93}.

The eikonal approximation is also valid in relativistic collisions, as it can
be derived from relativistic wave equations, e.g., the Klein-Gordon equation
\cite{Pil79}. Moreover, since it involves directions transverse to the beam,
it is relativistically invariant. The eikonal wavefunction also allows the
interpretation of the internal wave function variable $b$ as an impact
parameter. Thus, one can use the concept of classical trajectories, to obtain
excitation amplitudes and use them as input to the scattering amplitudes.

At intermediate energy collisions ($E_{lab} \simeq50$ MeV/nucleon), one must
perform a correction due to the Coulomb deflection of the particle's
trajectory. This correction amounts to calculating all elastic and inelastic
integrals replacing the asymptotic impact parameter $b$ by the distance of
closest approach in Rutherford orbits, i.e.,
\begin{equation}
b^{\prime}=a_{0}+\sqrt{a_{0}^{2}+b^{2}} \ ,\label{intro10}%
\end{equation}
where $a_{0}=Z_{1}Z_{2}e^{2}/m\mathrm{v}^{2}$ is half the distance of closest
approach in a head-on collision of point charged particles. This correction
leads to a considerable improvement of the eikonal amplitudes for the
scattering of heavy systems in collisions at intermediate energies.

As has been shown in ref. \cite{AAB90}, the nucleus-nucleus elastic and
inelastic scattering at intermediate and high energies is corrected
appropriately for relativistic kinematics if one replaces the quantity $a_{0}$
in eq. \ref{intro10} by $a^{\prime}_{0}=Z_{1}Z_{2}e^{2}/\gamma m\mathrm{v}%
^{2}$, where $\gamma=(1-\mathrm{v}^{2}/c^{2})^{-1/2}$ is the Lorentz factor.

\section{The optical potentials}

Usually, the optical potentials used in DWBA calculations are described in
terms of Woods-Saxon functions, both for the real and for the imaginary part
of the potentials, i.e.,
\begin{equation}
U_{opt}=-V_{0} \ f(r,R_{r},a_{r}) - iW_{0} \ f(r,R_{i},a_{i}) \ , \label{opt0}%
\end{equation}
where $f(r,R,a)=1/\{1+\exp[(r-R)/a]\}$. The parameters entering these
potentials are fitted to reproduce the elastic scattering data \cite{Satchler}.

In nuclear collisions at intermediate and high energies ($E_{lab} \geq50$
MeV/nucleon) the elastic scattering data are very scarce. One has to resort to
folding models with effective interactions, at least as a guide for the
experimental analysis. Among these models, the M3Y interaction is very
popular. It has been shown to work quite reasonably for elastic and inelastic
scattering of heavy ions at low and intermediate energy collisions
\cite{Ber77,Kob84}.

In its simplest form the M3Y interaction is given by two direct terms with
different ranges, and an exchange term represented by a delta interaction:
\begin{equation}
t(s)=A{\frac{e^{-\beta_{1} s}}{\beta_{1} s}}+ B{\frac{e^{-\beta_{2} s}}%
{\beta_{2} s}}+C\delta(\mathbf{s}) \ , \label{M3Y}%
\end{equation}
where A=7999 MeV, $B=-2134$ MeV, $C=-276$ MeV $fm^{3}$, $\beta_{1}=4$
$fm^{-1}$, and $\beta_{2}=2.5$ $fm^{-1}$. The real part of the optical
potential is obtained from a folding of this interaction with the ground state
densities, $\rho_{A}$ and $\rho_{B}$, of the nuclei A and B:
\begin{equation}
U_{M3Y}(\mathbf{r})=\int d^{3}r_{1}\; d^{3}r_{2} \; \rho_{A}(\mathbf{r}_{1})
\rho_{B}(\mathbf{r}_{2}) \; t(s) \ , \label{opt1}%
\end{equation}
with $\mathbf{s}=\mathbf{r}+\mathbf{r}_{2}-\mathbf{r}_{1}$. The imaginary part
of the optical potential is usually parameterized as Im $U_{opt}=\lambda
U_{M3Y}$, with $\lambda= 0.6-0.8$ \cite{Ber77,Kob84}.

This double folding M3Y potential yields values at the central region,
$r\simeq0$, which are too large compared to usual optical potentials. However,
one has to consider that the nuclear scattering at intermediate and high
energies ($E_{lab} \geq50$ MeV/nucleon) is mostly peripheral. Central
collisions will lead to fragmentation reactions which are not being considered
here. Thus, the only requirement here is that the optical potential reproduces
well the peripheral processes. This is the case of the M3Y potential and the
``t-$\rho\rho$" potential discussed below.

Another simple method to relate the nuclear optical potential to the
ground-state densities is the \textquotedblleft t$\rho\rho$" approximation.
This approximation has been extensively discussed in the literature
\cite{HRB91,Feshbach}. In its simplest version, neglecting the spin-orbit and
surface terms, the optical potential for proton-nucleus collisions is given
by
\begin{equation}
U_{opt}(\mathbf{r})=<t_{pn}>\rho_{n}(\mathbf{r})+<t_{pp}>\rho_{p}(\mathbf{r})
\end{equation}
where $\rho_{n}$ ($\rho_{p}$) are the neutron (proton) ground state densities
and $<t_{pi}>$ is the (isospin averaged) transition matrix element for
nucleon-nucleon scattering at forward directions,
\begin{equation}
t_{pi}(\mathbf{q}=0)=-{\frac{2\pi\hbar^{2}}{\mu}}f_{pi}(\mathbf{q}%
=0)=-{\frac{\hbar\mathrm{v}}{2}}\ \sigma_{pi}\ (\alpha_{pi}+i)
\end{equation}
where $\sigma_{pi}$ is the free proton-nucleon cross section and $\alpha_{pi}$
is the ratio between the imaginary and the real part of the proton-nucleon
scattering amplitude. The basic assumption here is that the scattering is
given solely in terms of the forward proton-nucleon scattering amplitude and
the local one-body density \cite{HRB91}.

For nucleus-nucleus collisions, the extension of this method leads to an
optical potential of the form
\begin{equation}
U_{opt}(\mathbf{r})=\int<t_{NN}(\mathbf{q}=0)> \ \rho_{A} (\mathbf{r-r^{\prime
}}) \ \rho_{B} (\mathbf{r^{\prime}}) \ d^{3}r^{\prime}\ , \label{opt4}%
\end{equation}
where $\mathbf{r}$ is the distance between the center-of-mass of the nuclei.
In this expression one uses the isospin average
\begin{equation}
<t_{NN}>={\frac{Z_{1}Z_{2}+N_{1}N_{2}}{A_{1}A_{2}}}t_{pp}+ {\frac{Z_{1}%
N_{2}+Z_{2}N_{1}}{A_{1}A_{2}}}t_{pn}.
\end{equation}

The parameters of the nucleon-nucleon cross scattering amplitudes for $E_{lab}
\geq100$ MeV/nucleon are shown in Table II, extracted from ref. \cite{Ra79}.
At lower energies one can use the isospin average values of Table II, which
describes well the nucleus-nucleus elastic scattering at lower energies
\cite{LVZ88}.

\begin{center}%
\begin{tabular}
[c]{|l|l|l|l|l|l|l|}\hline\hline
E & $\sigma_{pp}$ & $\alpha_{pp}$ & $\xi_{pp}$ & $\sigma_{pn}$ & $\alpha_{pn}$
& $\xi_{pn}$\\\hline
\lbrack MeV/nucl] & [mb] &  & [fm$^{2}$] & [mb] &  & [fm$^{2}$]\\\hline
100 & 33.2 & 1.87 & 0.66 & 72.7 & 1.00 & 0.36\\\hline
150 & 26.7 & 1.53 & 0.57 & 50.2 & 0.96 & 0.58\\\hline
200 & 23.6 & 1.15 & 0.56 & 42.0 & 0.71 & 0.68\\\hline
325 & 24.5 & 0.45 & 0.26 & 36.1 & 0.16 & 0.36\\\hline
425 & 27.4 & 0.47 & 0.21 & 33.2 & 0.25 & 0.27\\\hline
550 & 36.9 & 0.32 & 0.04 & 35.5 & -0.24 & 0.085\\\hline
650 & 42.3 & 0.16 & 0.07 & 37.7 & -0.35 & 0.09\\\hline
800 & 47.3 & 0.06 & 0.09 & 37.9 & -0.20 & 0.12\\\hline
1000 & 47.2 & -0.09 & 0.09 & 39.2 & -0.46 & 0.12\\\hline
2200 & 44.7 & -0.17 & 0.12 & 42.0 & -0.50 & 0.14\\\hline
\end{tabular}

Table I. Parameters \cite{Ra79} for the nucleon-nucleon amplitude, as given by
eq. \ref{opt6}.

\end{center}

The formula \ref{opt4} can be improved to account for the scattering angle
dependence of the nucleon-nucleon amplitudes. A good parametrization
\cite{Ra79} for the nucleon-nucleon scattering amplitude is given by
\begin{equation}
f_{NN}(\mathbf{q})= {\frac{k_{NN}}{4\pi}} \sigma_{NN} \ (i+\alpha
_{NN})\ e^{-\xi_{NN}\mathbf{q}^{2}} . \label{opt6}%
\end{equation}

The nuclear scattering phase then becomes \cite{Gl59}
\begin{equation}
\chi_{N}(\mathbf{b})=\int\int d \mathbf{r} \ d\mathbf{r}^{\prime}\ \rho
_{1}(\mathbf{r}) \ \gamma_{NN}(|\mathbf{b-s-s^{\prime}}|) \ \rho
_{2}(\mathbf{r}^{\prime}) \label{opt5}%
\end{equation}
where the profile function $\gamma_{NN}(\mathbf{b})$ is defined in terms of
the two-dimensional Fourier transform of the elementary scattering amplitude
\begin{equation}
\gamma_{NN} (\mathbf{b})= {\frac{1 }{2\pi i k_{NN}}} \ \int\exp
\Big[-i\mathbf{q.b}\Big] \ f_{NN}(\mathbf{q}) \ d\mathbf{q} \ ,
\end{equation}
and $\mathbf{s}$, $\mathbf{s^{\prime}}$ are the projections of the coordinate
vectors $\mathbf{r}$, $\mathbf{r}^{\prime}$ of the nuclear densities on the
plane perpendicular to the $z$-axis (beam-axis). For spherically symmetric
ground-state densities eq. \ref{opt5} reduces to the expression
\begin{equation}
\chi_{N}(b)=\int_{0}^{\infty}dq \ q \ \tilde\rho_{A}(q) \ f_{NN}(q)
\ \tilde\rho_{B}(q) \ J_{0}(qb) \ ,
\end{equation}
where $\tilde\rho_{i}(q)$ are the Fourier transforms of the ground state densities.

\begin{center}%
\begin{tabular}
[c]{|l|l|l|}\hline\hline
E [MeV/nucl] & $<\sigma_{NN}>$ [fm$^{2}$] & $<\alpha_{NN}>$\\\hline
30 & 19.6 & 0.87\\\hline
38 & 14.6 & 0.89\\\hline
40 & 13.5 & 0.9\\\hline
49 & 10.4 & 0.94\\\hline
85 & 6.1 & 1\\\hline
\end{tabular}

Table II. Same as in table I, but for lower incident energies \cite{LVZ88}.
The values are averaged over pp and pn collisions. $<\xi_{N}>$ is taken as
zero at these energies.
\end{center}

The optical potential can also be obtained by using an inversion method of the
eikonal phases. These phases might be chosen to fit the experimental data. In
this approach, one uses the Abel transform \cite{Gl59}
\begin{equation}
U_{opt}(r)={\frac{\hbar\mathrm{v}}{i\pi r}}{\frac{d}{dr}}\int_{r}^{\infty
}{\frac{\chi(b)}{(b^{2}-r^{2})^{1/2}}}\ r\ dr\ . \label{opt9}%
\end{equation}
This procedure has been tested in ref. \cite{AV87} leading to effective
potentials which on the tail, where the process takes place, are very close to
those obtained with phenomenological potentials of refs. \cite{Al84,Bue83}. It
can also be proved that under certain approximations, and for Gaussian density
distributions, the potential obtained through eq. \ref{opt9} coincides with
that obtained with the double folding procedure \cite{AV87}.

\section{Elastic scattering}

The elastic scattering in nucleus-nucleus collisions is a well established
tool for the investigation of ground state densities \cite{Satchler}. This is
because the optical potential can be related to the ground state densities by
means of a folding of the nucleon-nucleon interaction with the nuclear
densities of two colliding nuclei. But, as we have seen in the last section,
this relationship is not straightforward. It depends on the effective
interaction used, a proper treatment of polarization effects, and so on (for a
review see, e.g., \cite{HRB91}). At higher bombarding energies ($E_{Lab}
\geq50$ MeV/nucleon), a direct relationship between the nuclear densities and
the optical potential is possible, as long as the effects of multiple
nucleon-nucleon scattering can be neglected \cite{KMT}. The effects of real,
or virtual, nuclear excitations should also be considered, especially for
radioactive beams, involving small excitation energies.

The calculation of elastic scattering amplitudes using eikonal wavefunctions,
eq. \ref{intro2}, is very simple. They are given by \cite{Gl59}
\begin{equation}
f_{el}(\theta)=ik\ \int_{0}^{\infty}db\ b\ J_{0}(qb)\ \Big\{1-\exp
\Big[i\chi(b)\Big]\Big\}\ , \label{elast3}%
\end{equation}
where $q=2k\sin(\theta/2)$, and $\theta$ is the scattering angle. The elastic
scattering cross section is
\begin{equation}
{\frac{d\sigma_{el}}{d\Omega}}=\Big|f_{el}(\theta)\Big|^{2}\ . \label{elast3b}%
\end{equation}
For numerical purposes, it is convenient to make use of the analytical formula
for the Coulomb scattering amplitude. Thus, if one adds and subtracts the
Coulomb amplitude, $f_{C}(\theta)$ in eq. \ref{elast3}, one gets
\begin{equation}
f_{el}(\theta)=f_{C}(\theta)+ik\ \int_{0}^{\infty}db\ b\ J_{0}(qb)\ \exp
\Big[i\chi_{C}(b)\Big]\Big\{1-\exp\Big[i\chi_{N}(b^{\prime}%
)\Big]\Big\}\ ,\label{simplif}%
\end{equation}
where we replaced $b$ in $\chi_{N}(b)$ by $b^{\prime}$\ as given by eq.
\ref{intro10} to account for the nuclear recoil, as explained at the end of
section II.

The advantage in using this formula is that the term
$1-\exp\Big[i\chi _{N}(b)\Big]$ becomes zero for impact parameters
larger than the sum of the nuclear radii (grazing impact
parameter). Thus, the integral needs to be performed only within a
small range. In this formula, $\chi_{C}$ is given by eq.
\ref{intro4} and $f_{C}(\theta)$ is given by eq. \ref{fctheta},
with
\begin{equation}
\phi_{0}=-\eta C+\sum_{j=0}^{\infty}\left(  {\frac{\eta}{j+1}}-\arctan
{\frac{\eta}{j+1}}\right)  \ , \label{elast6}%
\end{equation}
and $C=0.5772156...$ is the Euler's constant.

At high energies the elastic scattering cross section for proton-nucleus
collisions is also well described by means of the eikonal approximation
\cite{Gl59}. The optical potential for proton-nucleus scattering is assumed to
be of the form \cite{Satchler}
\begin{equation}
U_{opt}(r)=U_{0}(r)+U_{S}(r)\ (\mathbf{L.S})+U_{C}(r)
\end{equation}
where
\begin{equation}
U_{0}(r)=V_{0}\ f_{R}(r)+iW_{0}\ f_{I}(r)+4\ i\ a_{S}\ W_{S}\ {\frac{d}{dr}%
}f_{I}(r) \label{U0pot}%
\end{equation}
and
\begin{equation}
U_{S}(r)=2\ \Big({\frac{\hbar}{m_{\pi}c}}\Big)^{2}\ V_{S}\ {\frac{1}{r}%
}\ {\frac{d}{dr}}f_{S}(r) \label{US}%
\end{equation}
are the central and spin-orbit part of the potential, respectively, and
$U_{C}(r)$ is the proton-nucleus Coulomb potential. The Fermi (or Woods-Saxon)
functions $f_{i}(r)$ are defined as before. The third term in eq. \ref{U0pot}
accounts for an increase of probability for nucleon-nucleon collisions at the
nuclear surface due to the Pauli principle. The existence of collective
surface modes, the possibility of nucleon transfer between ions suffering
peripheral collisions, and the possibility of breakup of the projectile and
the target make possible enhanced absorption at the surface. The spin-orbit
interaction in eq. \ref{US} is usually parametrized in terms of the pion mass,
$m_{\pi}$: $2(\hbar/m_{\pi}c)^{2}=4$ fm$^{2}$.

In the eikonal approximation, the proton-nucleus elastic scattering cross
section is given by \cite{Gl59}
\begin{equation}
{\frac{d\sigma_{el}}{d\Omega}}=\Big|F(\theta)\Big|^{2}+\Big|G(\theta
)\Big|^{2}\ ,
\end{equation}
where
\begin{equation}
F(\theta)=f_{C}(\theta)+ik\ \int_{0}^{\infty}db\ b\ J_{0}(qb)\ \exp
\Big[i\chi_{C}(b)\Big]\ \bigg\{1-\exp\Big[i\chi(b)\Big]\ \cos\Big[kb\ \chi
_{S}(b)\Big]\bigg\}
\end{equation}
and
\begin{equation}
G(\theta)=ik\ \int_{0}^{\infty}db\ b\ J_{1}(qb)\ \exp\Big[i\chi_{C}%
(b)+i\chi(b)\Big]\ \sin\Big[kb\ \chi_{S}(b)\Big]\ .
\end{equation}
In the equation above $q=2k\sin(\theta/2)$, where $\theta$ is the scattering
angle, $\chi=\chi_{N}+\chi_{C}$, $J_{0}$ ($J_{1}$) is the zero (first) order
Bessel function. The eikonal phase $\chi_{S}$ is given by \cite{Gl59}
\begin{equation}
\chi_{S}(\mathbf{b})=-{\frac{1}{\hbar\mathrm{v}}}\ \int_{-\infty}^{\infty
}U_{S}(\mathbf{b},\ z)\ dz\ . \label{elast20}%
\end{equation}

Eqs. \ref{elast3b}-\ref{elast20} describe the elastic scattering cross section
of $A\ (\mathrm{projectile}) + B$ in the center of mass system. In the
laboratory the scattering angle is given by \cite{Go80}
\begin{equation}
\theta_{L}=\arctan\left\{  {\frac{\sin\theta}{\gamma\left[  \cos\theta+ \rho
g(\rho, \mathcal{E}_{1}) \right]  }}\right\}  \ , \label{transf1}%
\end{equation}
where, $\rho=M_{A}/M_{B}$,
\begin{equation}
\mathcal{E}_{1}={\frac{E_{lab}\ \mathrm{[MeV/nucleon]} }{m_{N}c^{2}}} \ ,
\label{transf2}%
\end{equation}
where $m_{N}$ is the nucleon mass, and
\begin{equation}
g(\rho, \mathcal{E}_{1}) = {\frac{1+\rho(1+\mathcal{E}_{1}) }{1+\mathcal{E}%
_{1}+\rho}} \ , \ \ \ \ \ \gamma= {\frac{1+\mathcal{E}_{1} + \rho}%
{\sqrt{(1+\rho)^{2} +2\rho\mathcal{E}_{1}}}} \ . \label{transf3}%
\end{equation}
$\gamma$ is the relativistic Lorentz factor of the motion of the center of
mass system with respect to the laboratory.

The laboratory cross section is
\begin{equation}
{\frac{d\sigma_{el} }{d\Omega_{L}}} (\theta_{L}) = {\frac{ \left\{  \gamma^{2}
\left[  \rho g(\rho, \mathcal{E}_{1})+ \cos\theta\right]  ^{2} + \sin
^{2}\theta\right\}  ^{3/2} }{\gamma\left[  1+\rho g(\rho, \mathcal{E}_{1}%
)\cos\theta\right]  }} \ {\frac{d\sigma_{el} }{d\Omega}} (\theta)\ .
\label{transf4}%
\end{equation}

\section{Total nuclear reaction cross sections}

The total nuclear inelastic cross section (including fragmentation processes)
can be easily calculated within the optical limit of the Glauber model
\cite{Gl59}. It is given by
\begin{equation}
\sigma_{R}=2\pi\int_{0}^{\infty}\left[  1-T(b) \right]  b \ db \ ,
\end{equation}
where $T(b)$, the ``transparency function", is given by
\begin{equation}
T(b) = \exp\left[  2\mathrm{Im}\chi_{N} (b) \right]  \ .
\end{equation}

\section{The semiclassical method and coupled-channels problem}

Coulomb Excitation (CE) in high energy collisions is a well established tool
to probe several aspects of nuclear structure \cite{WA79,baur,Glasm}. The CE
induced by large-Z projectiles and/or targets, often yields large excitation
probabilities in grazing collisions. This results from the large nuclear
response to the acting electromagnetic fields. As a consequence, a strong
coupling between the excited states is expected.

Since there will be very little deflection by the Coulomb field in collisions
with impact parameter greater than the grazing one, the excitation amplitudes
can be calculated assuming a straight-line trajectory for the projectile. A
small Coulomb deflection correction can be used at the end, with the recipe
given by eq. \ref{intro10}.

We describe next a method for the calculation of multiple excitation among a
finite number of nuclear states. The system of coupled differential equations
for the time-dependent amplitudes of the eigenstates of the free nucleus is
solved numerically for electric dipole (E1), electric quadrupole (E2),
electric octupole (E3), magnetic dipole (M1), and magnetic quadrupole (M2) excitations.

Similarly, one can also calculate the amplitudes for (nuclear) monopole,
dipole, quadrupole, and octupole excitations.

In high energy nuclear collisions, the wavelength associated to
the projectile-target relative motion is much smaller than the
characteristic lengths of the system. It is, therefore, a
reasonable approximation to treat $\mathbf{r}$ as a classical
variable $\mathbf{r}(t)$, given at each instant by the trajectory
followed by the relative motion. At high energies it is also a
good approximation to replace this trajectory by a straight line.
The intrinsic dynamics can then be handled as a quantum mechanics
problem with a time-dependent Hamiltonian. This treatment is
discussed in full details by Alder and Winther in
ref.~\cite{AW65}.

The intrinsic state $|\psi(t)>$ satisfies the Schr\"{o}dinger equation
\begin{equation}
\left\{  H_{0}\ +\ V\left[  \mathbf{r}{(t)}\right]  \right\}  \mid
\psi(t)\rangle=i\hbar{\frac{\partial\mid\psi(t)\rangle}{\partial t}}\;.
\label{eqS}%
\end{equation}
Above, $H_{0}$ is the intrinsic Hamiltonian and $V$ is the channel-coupling interaction.

Expanding the wave function in the set $\{\mid j\rangle;\ j=1,N\}$ of
eigenstates of $H_{0}$, where $N$ is the number states included in the
coupled-channels (CC) problem, we obtain a set of coupled equations. Taking
the scalar product with each of the states $<k|$, we get
\begin{equation}
i\hbar\ {\dot{a}}_{k}(t)=\sum_{j=1}^{N}\ \langle k\mid V(t)\mid j\rangle
\;\exp\left[  i(E_{k}-E_{j})t/\hbar\right]  \;a_{j}(t)\;,\qquad\qquad
k=1\;\mathrm{to}\;\;N\ , \label{AW}%
\end{equation}
where $E_{n}$ is the energy of the state $\left|  n\right\rangle .$ It should
be remarked that the amplitudes depend also on the impact parameter $b$
specifying the classical trajectory followed by the system. For the sake of
keeping the notation simple, we do not indicate this dependence explicitly. We
write, therefore, $a_{n}(t)$ instead of $a_{n}(b,t)$, restoring the notation
with $b$, or $t$, whenever necessary. Since the interaction $V $ vanishes as
$t\rightarrow\pm\infty$, the amplitudes have as initial condition
$a_{n}(t\rightarrow-\infty)=\delta_{n1}$ and they tend to constant values as
$t\rightarrow\infty$.

A convenient measure of time is given by the dimensionless quantity
$\tau=\gamma\mathrm{v}t/b$, where $\gamma=(1-\mathrm{v}^{2}/c^{2})^{-1/2}$ is
the Lorentz factor for the projectile velocity v. A convenient measure of
energy is $E_{0}=\gamma\hbar\mathrm{v}/b.$ In terms of these quantities the CC
equations become
\begin{equation}
\frac{da_{k}(\tau)}{d\tau}=-i\sum_{j=1}^{N}\ \langle k\mid W(\tau)\mid
j\rangle\;\exp\left(  i\xi_{kj}\tau\right)  \;a_{j}(\tau)\;;\;\;\;W(\tau
)=\frac{V(\tau)}{E_{0}}\;;\;\;\xi_{kj}=\frac{E_{k}-E_{j}}{E_{0}}\;\;.
\label{AW2}%
\end{equation}

The nuclear states are specified by the spin quantum numbers $I$ and $M$.
Therefore, the excitation probability of an intrinsic state $\mid
n\rangle\equiv\mid I_{n},M_{n}\rangle$ in a collision with impact parameter
$b$ is obtained from an average over the initial orientation $\left(
M_{1}\right)  ,$ and a sum over the final orientation of the nucleus,
respectively:
\begin{equation}
P_{n}(b)=\frac1{2I_{1}+1}\sum_{M_{1},M_{n}}|a_{I_{n},M_{n}}^{M_{1}}(b)|^{2}\;.
\label{Pn}%
\end{equation}
The total cross section for excitation of the state $|n>$ is obtained by the
classical expression
\begin{equation}
\sigma_{n}=2\pi\ \int\ P_{n}(b)\ b\;db\;.
\end{equation}

\section{Time-dependent electromagnetic interaction}

We consider a nucleus 2 which is at rest and a projectile nucleus 1 which
moves along the $z$-axis. Nucleus 2 is excited from the initial state
$|I_{j}M_{j}>$ to the state $|I_{k}M_{k}>$ by the electromagnetic field of
nucleus 1. The nuclear states are specified by the spin quantum numbers
$I_{j}$, $I_{k}$ and by the corresponding magnetic quantum numbers $M_{j}$ and
$M_{k}$. We assume that nucleus 1 moves along a straight-line trajectory with
impact parameter $b$, which is therefore also the distance of the closest
approach between the center of mass of the two nuclei at the time $t=0$. The
interaction, $V_{C}(t),$ due to the electromagnetic field of the nucleus 1
acting on the charges and currents nucleus 2 can be expanded into multipoles,
as explained in ref. \cite{BC96}. One has
\begin{equation}
W_{C}(\tau)=\frac{V_{C}(\tau)}{E_{0}}=\sum_{\pi\lambda\mu}W_{\pi\lambda\mu
}(\tau)\ , \label{Vfi3}%
\end{equation}
where $\pi=E,\;M$ denotes electric and magnetic interactions, respectively,
and
\begin{equation}
W_{\pi\lambda\mu}(\tau)=\left(  -1\right)  ^{\lambda+1}\frac{Z_{1}e}%
{\hbar\mathrm{v}b^{\lambda}}\frac{1}{\lambda}\;\sqrt{\frac{2\pi}{\left(
2\lambda+1\right)  !!}}\;Q_{\pi\lambda\mu}(\xi,\tau)\;\mathcal{M}(\pi
\lambda,-\mu)\;, \label{wlamb}%
\end{equation}
where $\mathcal{M}(\pi\lambda,\mu)$ is the multipole moment of order
$\lambda\mu$ \cite{EG88},
\begin{equation}
\mathcal{M}(E\lambda,\mu)=\int d^{3}r\ \rho_{C}(\mathbf{r})\ r^{\lambda
}\ Y_{1\mu}(\mathbf{r})\ , \label{ME1}%
\end{equation}
and
\begin{equation}
\mathcal{M}(M1,\mu)=-{\frac{i}{2c}}\ \int d^{3}r\ \mathbf{J}_{C}%
(\mathbf{r}).\mathbf{L}\left(  rY_{1\mu}\right)  \ , \label{MM1}%
\end{equation}
$\rho_{C}$ ($\mathbf{J}_{C}$) being the nuclear charge (current). The
quantities $Q_{\pi\lambda\mu}(\tau)$ were calculated in ref. \cite{BC96,Ber99}%
, and for the E1, E2, and M1 multipolarities. We will extend the formalism to
M2 and E3 excitations.

We use here the notation of Edmonds \cite{Ed60} where the reduced multipole
matrix element is defined by
\begin{equation}
\mathcal{M}_{kj}(\pi\lambda,\mu)=(-1)^{I_{k}-M_{k}}\ \left(  {{{{{\ { {%
\genfrac{}{}{0pt}{}{I_{k} }{-M_{k}}%
} } }}}}}{{{{{\ { {%
\genfrac{}{}{0pt}{}{\lambda}{\mu}%
} } }}}}}{{{{{{ {%
\genfrac{}{}{0pt}{}{I_{j} }{M_{j}}%
} }}}}}}\right)  <I_{k}||\mathcal{M}(\pi\lambda)||I_{j}>\ . \label{Mfi1}%
\end{equation}

To simplify the expression (\ref{AW2}) we introduce the dimensionless
parameter $\psi_{kj}^{(\lambda\mu)}$ by the relation
\[
\psi_{kj}^{(\lambda\mu)}=\left(  -1\right)  ^{\lambda+1}\frac{Z_{1}e}%
{\hbar\mathrm{v}b^{\lambda}}\frac{1}{\lambda}\sqrt{\frac{2\pi}{\left(
2\lambda+1\right)  !!}}\;\mathcal{M}_{kj}(\pi\lambda, -\mu) \ .
\]
Then we may write eq. \ref{AW2} in the form
\begin{equation}
\frac{da_{k}(\tau)}{d\tau}=-i\sum_{r=1}^{N}\ \sum_{\pi\lambda\mu}Q_{\pi
\lambda\mu}(\xi_{kj},\tau)\psi_{kj}^{\left(  \lambda\mu\right)  }\;\exp\left(
i\xi_{kj}\tau\right)  \;a_{j}(\tau)\;. \label{ats1}%
\end{equation}

The explicit expressions for eq. \ref{wlamb} can also be obtained by a Fourier
transform of the excitation amplitudes found in ref. \cite{WA79}, i.e.,
\begin{equation}
W_{\pi\lambda\mu}(\tau)=\frac{1}{E_{0}}\cdot\frac{1}{2\pi}\int_{-\infty
}^{\infty}e^{i\omega t}\;V_{\pi\lambda\mu}(\omega)d\omega=\frac{1}{2\pi\hbar
}\int_{-\infty}^{\infty}e^{i\xi\tau}\;V_{\pi\lambda\mu}(\xi)d\xi,
\label{Fourier1}%
\end{equation}
where $\omega=(E_{k}-E_{j})/\hbar=E_{0}\xi/\hbar$ \ (here we omit the
sub-indexes $kj$ for convenience). The expressions for $V_{\pi\lambda\mu
}(\omega_{ij})$ are given by \cite{WA79}
\begin{equation}
V_{\pi\lambda\mu}(\omega_{ij})=\frac{Z_{1}e}{\mathrm{v}\gamma}\left(
-1\right)  ^{\mu}\sqrt{2\lambda+1}\left(  \frac{\omega_{ij}}{c}\right)
^{\lambda}G_{\pi\lambda\mu}\left(  \frac{c}{\mathrm{v}}\right)  K_{\mu}\left(
\xi_{ij}\right)  \mathcal{M}(\pi\lambda,-\mu). \label{WApot}%
\end{equation}

Using the properties of $V_{\pi\lambda\mu}(\omega)$ for negative $\omega$, one
can show that $V_{\pi\lambda\mu}(-\omega)=(-1)^{\lambda+\mu}V_{\pi\lambda\mu
}^{\ast}(\omega)$. Then, one gets from eq. \ref{Fourier1}
\begin{equation}
W_{\pi\lambda\mu}(\tau)=\frac{Z_{1}e}{2\pi\hbar\mathrm{v}\gamma}\left(
-1\right)  ^{\mu}\left(  {\frac{\gamma\mathrm{v}}{bc}}\right)  ^{\lambda
}\;\sqrt{2\lambda+1}\;G_{\pi\lambda\mu}\left(  \frac{c}{\mathrm{v}}\right)
F_{\lambda\mu}\left(  \xi_{ij},\tau\right)  \mathcal{M}(\pi\lambda,-\mu),
\label{WApotw}%
\end{equation}
where
\begin{align}
F_{\lambda\mu}\left(  \xi,\tau\right)   &  =2\int_{0}^{\infty}\cos\left(
\xi\tau\right)  \xi^{\lambda}K_{\mu}\left(  \xi\right)  d\xi
,\;\;\;\;\;\mathrm{for}\;\;\;\;\;\lambda+\mu=\mathrm{even},\nonumber\\
&  =\frac{2}{i}\int_{0}^{\infty}\sin\left(  \xi\tau\right)  \xi^{\lambda
}K_{\mu}\left(  \xi\right)  d\xi,\;\;\;\;\;\mathrm{for}\;\;\;\;\;\lambda
+\mu=\mathrm{odd}. \label{Flm}%
\end{align}
\bigskip These integrals can be obtained analytically.

Using the functions $G_{\pi\lambda\mu}\left(  c/\mathrm{v}\right)  $ derived
in ref. \cite{WA79} we get explicit closed forms for the E1, E2, E3, M1 and M2
multipolarities,
\begin{equation}
Q_{E10}(\xi_{ij},\tau)=-\frac{\sqrt{2}}{\gamma}\tau\phi^{3}(\tau
)\;;\;\;\;\;\;Q_{E1\pm1}(\xi_{ij},\tau)=\mp\phi^{3}(\tau)\;, \label{QE1}%
\end{equation}%
\begin{equation}
Q_{M10}(\xi_{ij},\tau)=0\;;\;\;\;\;\;Q_{M1\pm1}(\xi_{ij},\tau)=i\left(
\frac{\mathrm{v}}{c}\right)  \phi^{3}(\tau)\;, \label{QM1}%
\end{equation}%
\begin{align}
Q_{E20}(\xi_{ij},\tau)  &  =\sqrt{6}\left(  2\tau^{2}-1\right)  \phi^{5}%
(\tau)\;,\nonumber\\
Q_{E2\pm1}(\xi_{ij},\tau)  &  =\pm3\left(  2-\beta^{2}\right)  \gamma\tau
\phi^{5}(\tau)\;;\;\;\;\;Q_{E2\pm2}(\tau)=3\phi^{5}(\tau)\;, \label{QE2}%
\end{align}%
\begin{align}
Q_{M20}(\xi_{ij},\tau)  &  =0\;,\nonumber\\
Q_{M2\pm1}(\xi_{ij},\tau)  &  =-3i\left(  \frac{\mathrm{v}}{c}\right)
\gamma\tau\phi^{5}(\tau)\;;\;\;\;\;Q_{M2\pm2}(\tau)=\mp3i\left(
\frac{\mathrm{v}}{c}\right)  \phi^{5}(\tau)\;, \label{QM2}%
\end{align}
and
\begin{align}
Q_{E30}(\xi_{ij},\tau)  &  =-3\sqrt{\frac{3}{10}}\gamma\left(  5-\beta
^{2}\right)  \tau\left(  2\tau^{2}-3\right)  \phi^{7}(\tau)\;,\nonumber\\
Q_{E3\pm1}(\xi_{ij},\tau)  &  =\mp\frac{3}{2\sqrt{10}}\left(  15-11\beta
^{2}\right)  \gamma^{2}\left(  4\tau^{2}-1\right)  \phi^{7}(\tau
)\;,\nonumber\\
Q_{E3\pm2}(\xi_{ij},\tau)  &  =-\gamma\frac{15}{2}\left(  3-\beta^{2}\right)
\tau\phi^{7}(\tau);\;\;\;\;Q_{E3\pm3}(\xi_{ij},\tau)=\mp\frac{15}{2}%
\sqrt{\frac{3}{2}}\phi^{7}(\tau)\;, \label{QE3}%
\end{align}
where $\phi\left(  \tau\right)  =\left(  1+\tau^{2}\right)  ^{-1/2}$, and
$\beta=\mathrm{v}/c$.

The fields $Q_{\pi\lambda\mu}(\xi_{ij},\tau)$ peak around $\tau=0$, and
decrease rapidly within an interval $\Delta\tau\simeq1$, corresponding to a
collision time $\Delta t\simeq b/\gamma\mathrm{v}$. This means that
numerically one needs to integrate the CC equations in time within an interval
of range $n\times\Delta\tau$ around $\tau=0$, with $n$ equal to a small
integer number.

It is important to notice that the multipole interactions derived in this
section assume that the Coulomb potential has an $1/r$ shape (when
$\gamma\rightarrow1$) even inside the nuclei. This is a simplification
justified by the strong absorption at small impact parameters for which the
Coulomb potential deviates from its $1/r$ form. We have not included the
deviations to the $1/r$ behavior of the Coulomb interaction inside the nuclei,
as the proper relativistic treatment of it is rather complicated. This has
been discussed in details in ref. \cite{Esb02}.

\section{Time-dependent nuclear excitation: Collective model}

In peripheral collisions the nuclear interaction between the ions can also
induce excitations. According to the collective, or Bohr-Mottelson,
particle-vibrator coupling model the matrix element for the transition
$j\longrightarrow k$ is given by \cite{Satchler,Sa87}
\begin{equation}
V_{N(\lambda\mu)}^{(kj)}(\mathbf{r})\equiv<I_{k}M_{k}|V_{N(\lambda\mu)}%
|I_{j}M_{j}>=-{\frac{\delta_{\lambda}}{\sqrt{2\lambda+1}}}\ <I_{k}%
M_{k}|Y_{\lambda\mu}|I_{j}M_{j}>\ Y_{\lambda\mu}(\hat{\mathbf{r}}%
)\ U_{\lambda}(r) \label{VfiN}%
\end{equation}
where $\delta_{\lambda}$ is the vibrational amplitude and $U_{\lambda}(r)$ is
the transition potential. To follow the convention of ref. \cite{Sa87}, we use
$\alpha_{0}$ instead of $\delta_{0}$ in the equation above.

The transition potentials for nuclear excitations can be related to the
optical potential in the elastic channel. This is discussed in details in ref.
\cite{Sa87}. The transition potentials for isoscalar excitations are
\begin{equation}
U_{0}(r)=3U_{opt}(r)+r{\frac{dU_{opt}(r)}{dr}}\ , \label{U0}%
\end{equation}
for monopole,
\begin{equation}
U_{1}(r)={\frac{dU_{opt}}{dr}}+{\frac13}\ R_{0}\ {\frac{d^{2}U_{opt}}{dr^{2}}%
}\ , \label{U1}%
\end{equation}
for dipole, and%

\begin{equation}
U_{2}(r)={\frac{dU_{opt}(r)}{dr}}\ , \label{U2}%
\end{equation}
for quadrupole and octupole modes. $R_{0}$ is the nuclear radius at ${\frac
{1}{2}}$ the central nuclear density.

The deformation length $\delta_{\lambda}$ can be directly related to the
reduced matrix elements for electromagnetic transitions. Using well-known
sum-rules for these matrix elements one finds a relation between the
deformation length, and the nuclear sizes and the excitation energies. For
isoscalar excitations one obtains \cite{Sa87}
\begin{equation}
\alpha_{0}^{2}= 2 \pi\ {\frac{\hbar^{2} }{m_{N}}} \ {\frac{1}{<r^{2}> A E_{x}%
}} \ , \ \ \ \ \ \ \ \ \ \delta_{\lambda\geq2}^{2} = {\frac{2 \pi}{3}}
\ {\frac{\hbar^{2} }{m_{N}}} \ \lambda\ (2\lambda+1) \ {\frac{1}{A E_{x}}}
\label{deform1}%
\end{equation}
where $A$ is the atomic number, $<r^{2}>$ is the r.m.s. radius of the nucleus,
and $E_{x}$ is the excitation energy.

For dipole isovector excitations \cite{Sa87}
\begin{equation}
\delta_{1}^{2}= {\frac{\pi}{2}} \ {\frac{\hbar^{2} }{m_{N}}} \ {\frac{A }{NZ}}
\ {\frac{1}{E_{x}}}\ ,
\end{equation}
where $Z$ ($N$) the charge (neutron) number. The transition potential in this
case is modified from eq. \ref{U1} to account for the isospin dependence
\cite{Sa87}. It is given by
\begin{equation}
U_{1}(r)=-\Lambda\ \Big( {\frac{N-Z }{A}} \Big) \ \Big( {\frac{dU_{opt} }{dr}}
+ {\frac{1}{3}} \ R_{0} \ {\frac{d^{2} U_{opt} }{dr^{2}}} \Big) \ ,
\end{equation}
where the factor $\Lambda$ depends on the difference between the proton and
the neutron matter radii as
\begin{equation}
\Lambda{\frac{2(N-Z)}{3A}} = {\frac{R_{n}-R_{p} }{{\frac{1}{2}} \ (R_{n}%
+R_{p})}} = {\frac{\Delta R_{np} }{R_{0}}} \ . \label{deform11}%
\end{equation}
Thus, the strength of isovector excitations increases with the difference
between the neutron and the proton matter radii. This difference is
accentuated for neutron-rich nuclei and should be a good test for the quantity
$\Delta R_{np}$ which enters the above equations.

Notice that the reduced transition probability for electromagnetic transitions
is defined by \cite{EG88}
\begin{equation}
B\left(  \pi\lambda; i\rightarrow j\right)  = {\frac{1}{2I_{i}+1}} \ \left|
<I_{j}||\mathcal{M}(\pi\lambda)||I_{i}> \right|  ^{2} \ . \label{Bvalue}%
\end{equation}
These can be related to the deformation parameters by \cite{Sa87}
\begin{equation}
B(E0) = \left[  {\frac{3 ZeR_{0}^{2}}{10\pi}} \right]  ^{2} \alpha_{0}^{2}\ ,
\ \ \ \ \ \ B(E1) = {\frac{9}{4\pi}}\left(  {\frac{NZe}{A}}\right)  ^{2}
\delta_{1}^{2} \ ,
\end{equation}
and
\begin{equation}
B(E\lambda)_{\lambda\geq2} = \left[  {\frac{3 }{4\pi}} ZeR_{0}^{\lambda
-1}\right]  ^{2} \delta_{\lambda}^{2} \ .
\end{equation}

The time dependence of the matrix elements above can be obtained by making a
Lorentz boost. One gets
\begin{align}
V_{N(\lambda\mu)}^{(kj)}(t)  &  \equiv<I_{k}M_{k}|U|I_{j}M_{j}>\nonumber\\
&  =-\gamma\ {\frac{\delta_{\lambda}}{\sqrt{2\lambda+1}}}\ <I_{k}%
M_{k}|Y_{\lambda\mu}|I_{j}M_{j}>Y_{\lambda\mu}\left(  \theta(t){,0}\right)
\ U_{\lambda}[r(t)]\ , \label{VfiN2}%
\end{align}
where $r(t)=\sqrt{b^{2}+\gamma^{2}\mathrm{v}^{2}t^{2}}=b/\phi\left(
\tau\right) $, $\theta=\tau\phi(\tau)$, and%

\begin{equation}
<I_{k}M_{k}|Y_{\lambda\mu}|I_{j}M_{j}>=(-1)^{I_{k}-M_{k}}\ \left[  {\frac{
(2I_{k}+1)(2\lambda+1)}{4\pi(2I_{j}+1)}}\right]  ^{1/2}\ \left(  {{{{{ {%
\genfrac{}{}{0pt}{}{I_{k} }{-M_{k}}%
} }}}}}{{{{{ {%
\genfrac{}{}{0pt}{}{\lambda}{\mu}%
} }}}}}{}{{{{{ {%
\genfrac{}{}{0pt}{}{I_{j} }{M_{j}}%
} }}}}}\right)  \left(  {{{{{ {%
\genfrac{}{}{0pt}{}{I_{k} }{0}%
} }}}}}{{{{{ {%
\genfrac{}{}{0pt}{}{\lambda}{0}%
} }}}}}{{{{{ {%
\genfrac{}{}{0pt}{}{I_{j} }{0}%
} }}}}}\right)  \ . \label{WE}%
\end{equation}

To put it in the same notation as in eq. (\ref{ats1}), we define
$Q_{N\lambda\mu}^{(kj)}(\tau)=V_{N(\lambda\mu)}^{(kj)}(t)/E_{0}$, and the
coupled-channels equations become%

\begin{equation}
\frac{da_{k}(\tau)}{d\tau}=-i\sum_{j=1}^{N}\ \sum_{\lambda\mu} \sum_{\pi
}\left[  Q_{N\lambda\mu}^{(kj)}(\xi_{kj},\tau)+Q_{C\pi\lambda\mu}(\xi
_{kj},\tau)\right]  \psi_{kj}^{\left(  \lambda\right)  } \;\exp\left(
i\xi_{kj}\tau\right)  \;a_{j}(\tau)\;. \label{ats2}%
\end{equation}

\section{Absorption at small impact parameters}

If the optical potential $U_{opt}(\mathbf{r})$ is known, the absorption
probability in grazing collisions can be calculated in the eikonal
approximation as%

\begin{equation}
A(b)=\exp\left[  \frac2{\hbar\mathrm{v}}\int_{-\infty}^{\infty}\mathrm{Im}%
\left[  U_{opt}(\mathbf{r})\right]  dz\right]  \;, \label{abs}%
\end{equation}
where $r=\sqrt{b^{2}+z^{2}}$. If the optical potential is not known, the
absorption probability can be calculated from the optical limit of the Glauber
theory of multiple scattering (also from the ``t-$\rho\rho$" approximation),
which yields:%

\begin{equation}
A(b)=\exp\left\{  -\sigma_{NN}\int_{-\infty}^{\infty}\ \left[  \int\rho
_{1}(\mathbf{r}^{\prime})\ \rho_{2}(\mathbf{r-r^{\prime}})\ d^{3}r^{\prime
}\right]  dz\right\}  \;. \label{abs2}%
\end{equation}
where $\sigma_{NN}$ is the nucleon-nucleon cross section and $\rho_{i}$ is the
ground state density of the nucleus $i.$ For stable nuclei, these densities
are taken from the droplet model densities of Myers and Swiatecki \cite{MS69},
but can be easily replaced by more realistic densities.

Including absorption, the total cross section for excitation of the state
$|n>$ is obtained by
\begin{equation}
\sigma_{n}=2\pi\ \int\ A(b)P_{n}(b)\ bdb\;. \label{sigman2}%
\end{equation}

\section{Angular distribution of inelastically scattered particles}

The angular distribution of the inelastically scattered particles can be
obtained from the semiclassical amplitudes, $a_{I_{n},M_{n}}^{M_{1}}(b)$,
described in section VII. For the excitation of a generic state $|n>$, it is
given by \cite{BN93}
\begin{equation}
f_{inel}^{\mu}(\theta)= ik\int_{0}^{\infty}db \ b \ J_{\mu}(qb) \ e^{i\chi(b)}
\ a_{\mu}(b) \ , \label{angp1}%
\end{equation}
where we simplified the notation: $a_{\mu}\equiv a_{I_{n},M_{n}}^{M_{1}}$,
with $\mu=M_{n} - M_{1}$.

The inelastic scattering cross section is obtained by an average over the
initial spin and a sum over the final spin:
\begin{equation}
{\frac{d\sigma_{inel}}{d\Omega}} = \frac1{2I_{1}+1} \sum_{M_{1},M_{n}}
|f_{inel}^{\mu}|^{2} \ . \label{angp91}%
\end{equation}

The program DWEIKO uses eq. \ref{angp91} to calculate the angular distribution
in inelastic scattering. But it is instructive to show how it relates to the
usual semiclassical approximation. For collisions at high energies, the
integrand of eq. \ref{angp1} oscillates wildly at the relevant impact
parameters and scattering angles. One can use the approximation
\begin{align}
J_{\mu}(qb)  &  \simeq\sqrt{\frac{2}{\pi qb}} \cos\left(  qb - {\frac{\pi\mu
}{2}} - {\frac{\pi}{4}} \right) \nonumber\\
&  = {\frac{1}{\sqrt{2\pi qb}}} \left\{  e^{iqb}e^{-i\pi(\mu+1/2)/2}%
+e^{-iqb}e^{i\pi(\mu+1/2)/2} \right\}  \ , \label{angp2}%
\end{align}
together with the stationary-phase approximation \cite{MF53}
\begin{equation}
\int G(x) e^{i\phi(x)} dx \simeq\left(  {\frac{2\pi i }{\phi^{\prime\prime
}(x_{0})}} \right)  ^{1/2} G(x_{0}) e^{i\phi(x_{0})} \ , \label{angp3}%
\end{equation}
where $x_{0}$ is the point of stationary phase, satisfying
\begin{equation}
\phi^{\prime}(x_{0})=0 \ . \label{angp4}%
\end{equation}
This approximation is valid for a slowly varying function $G(x)$.

Only the second term in the brackets of eq. \ref{angp2} will have a positive
($b=b_{0}>0$) stationary point, and eq. \ref{angp1} becomes
\begin{equation}
f_{inel}^{\mu}(\theta)\simeq i{\frac{k}{\sqrt{q}}} \left(  {\frac{i}%
{\phi^{\prime\prime}(x_{0})}}\right)  ^{1/2} \sqrt{b_{0}} \exp\left[
{\mathrm{Im} \chi_{N}(b_{0})}\right]  \ \exp\left[  i\chi(b_{0})+i\pi
(m+1/2)/2\right]  \ a_{\mu}(b_{0})\ , \label{angp5}%
\end{equation}
where
\begin{equation}
\phi=-qb+2\eta\ln(kb) + \mathrm{Re} \chi^{\prime}_{N}(b) \ , \label{angp6}%
\end{equation}
and $b_{0}$, the ``classical impact parameter" is the solution of
\begin{equation}
-q + {\frac{2\eta}{b_{0}}} + \mathrm{Re} \chi^{\prime}_{N} (b_{0}) = 0 \ .
\label{angp7}%
\end{equation}
This equation has 2 solutions: (a) one corresponding to \textit{close} (or
\textit{nearside}) collisions, (b) and another corresponding to \textit{far}
(or \textit{farside}) collisions. These are collisions passing by one side and
the opposite side of the target, but leading to the same scattering angle.
They thus lead to interferences in the cross sections.

In collisions at high energies, the inelastic scattering is dominated by close
collisions and, moreover, one can neglect the third term in eq. \ref{angp7}.
The condition $\phi^{\prime}(b_{0})=0$ implies
\begin{equation}
b_{0}={\frac{2\eta}{q}}={\frac{a_{0}}{\sin(\theta/2)}}\ ,\ \ \ \ a_{0}%
={\frac{Z_{1}Z_{2}e^{2}}{2k\mathrm{v}}}\ ,\ \ \ \ \mathrm{and}\ \ \phi
^{\prime\prime}(b_{0})=-{\frac{2\eta}{b_{0}^{2}}}=-{\frac{q^{2}}{2\eta}}\ .
\label{angp8}%
\end{equation}
We observe that the relation \ref{angp8} is the same [with $\cot
(\theta/2)\simeq\sin^{-1}(\theta/2)$] as that between the impact parameter and
the deflection angle of a particle following a classical Rutherford trajectory.

Inserting these results in eqs. \ref{angp1} and \ref{angp91}, one gets
\begin{equation}
{\frac{d\sigma_{inel}^{(n)}}{d\Omega}} = \left(  {\frac{4\eta^{2}k^{2}}{q^{4}%
}}\right)  \frac1{2I_{1}+1}\sum_{M_{1},M_{n}}|a_{I_{n},M_{n}}^{M_{1}}%
(b_{0})|^{2} \ e^{2\mathrm{Im} \chi_{N}(b_{0})} \ . \label{angp10}%
\end{equation}
One can easily see that the factor $4\eta^{2}k^{2}/ q^{4}$ is the Rutherford
cross section.

The above results show that the description of the inelastic scattering in
terms of the eikonal approximation reproduces the expected result, i.e., that
the excitation cross sections are determined by the product of the Rutherford
cross sections and the excitation probabilities. This is a commonly used
procedure in Coulomb excitation at low energies.

The cross sections in the laboratory system are obtained according to the same
prescription as described at the end of section IV.

\section{Angular distribution of $\gamma$-rays}

\begin{figure}[tb]
\begin{center}
\includegraphics[width=4.5in,angle=0,clip=true]{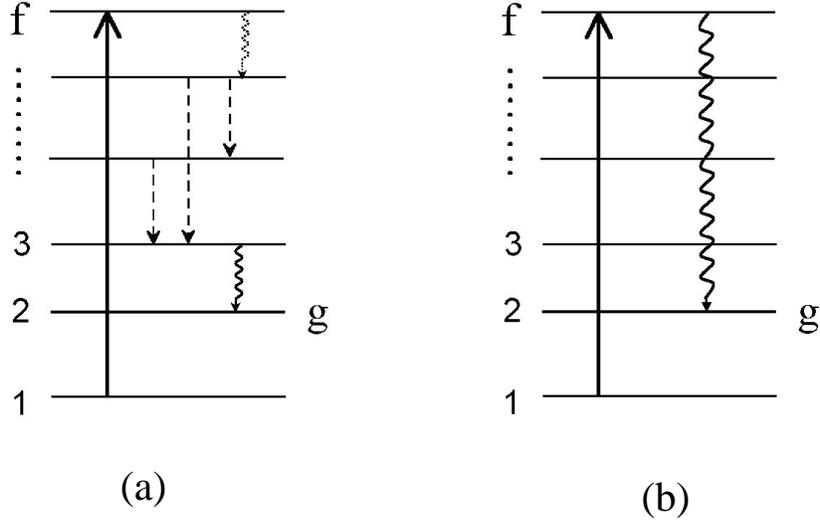}
\end{center}
\par
\vspace*{0pt}\caption{Schematic description of a nuclear excitation (solid
line) followed by $\gamma$-decay (solid wavy line). (a) The dashed lines are
transitions due to internal conversion (unobserved). The dashed wavy line is
an unobserved $\gamma$-decay. (b) Direct emission of an observed gamma ray. }%
\label{cascade1}%
\end{figure}

After the excitation, the nuclear state $\left|  I_{f}\right\rangle $ can
decay by gamma emission to another state $\left|  I_{g}\right\rangle $.
\ Complications arise from the fact that the nuclear levels are not only
populated by Coulomb excitation, but also by conversion and $\gamma
$-transitions cascading down from higher states (see figure \ref{cascade1}%
(a)). To compute the angular distributions one must know the parameters
$\Delta_{l}\left(  i\longrightarrow j\right)  $ and $\epsilon_{l}\left(
i\longrightarrow j\right)  $\, for $l\geq1$ \cite{AW65},
\begin{equation}
\epsilon_{l}^{2}\left(  i\longrightarrow j\right)  =\alpha_{l}\left(
i\longrightarrow j\right)  \Delta_{l}^{2}\left(  i\longrightarrow j\right)  ,
\label{casc1}%
\end{equation}
where $\alpha_{l}$ is the total $l$-pole conversion coefficient, and
\begin{equation}
\Delta_{\pi l}=\left[  \frac{8\pi\left(  l+1\right)  }{l\left[  \left(
2l+1\right)  !!\right]  ^{2}}\frac{1}{\hbar}\left(  \frac{\omega}{c}\right)
^{2l+1}\right]  ^{1/2}\left(  2I_{j}+1\right)  ^{-1/2}\left\langle
I_{j}\left\|  i^{s(l)}\mathcal{M}(\pi l)\right\|  I_{i}\right\rangle \ ,
\label{gad0_1}%
\end{equation}
with $s(l)=l$ for electric $\left(  \pi=E\right)  $ and $s(l)=l+1$ for
magnetic $\left(  \pi=M\right)  $ transitions. The square of $\Delta_{\pi l}$
is the $l$-pole $\gamma$-transition rate (in $\sec^{-1}$).

As for the non-relativistic case \cite{WB65,AW65}, the angular distributions
of gamma rays following the excitation depend on the frame of reference used.
In our notation, the z-axis corresponds to the beam axis, and the statistical
tensors are given by (we use the notation of \cite{WB65,AW65})
\begin{align}
\alpha_{k\kappa}^{(0)}(f)  &  =\frac{\left(  2I_{f}+1\right)  ^{1/2}}{\left(
2I_{1}+1\right)  }\sum_{M_{f}=-(M_{f}^{^{\prime}}+\kappa),M_{f}^{^{\prime}}%
}\left(  -1\right)  ^{I_{f}+M_{f}}\left(
\begin{tabular}
[c]{lll}%
$I$$_{f}$ & $I$$_{f}$ & $k$\\
$-M$$_{f}$ & $M$$_{f}^{^{\prime}}$ & $\kappa$%
\end{tabular}
\right) \nonumber\\
&  \times\sum_{M_{1}}a_{I_{f}M_{f}^{^{\prime}}}^{\ast}(M_{1})\;a_{I_{f}M_{f}%
}(M_{1})\;, \label{stat}%
\end{align}
where $f$ is the state from which the gamma ray is emitted, and $1$ denotes
the initial state of the nucleus, before the excitation. To calculate the
angular distributions of the gamma rays one needs the statistical tensors for
$k=0,2,4$ and $-k\leq\kappa\leq k$ (see \cite{WB65,AW65}).

Instead of the diagram of figure \ref{cascade1}(a), we will consider here the
much simpler situation in which the $\gamma$-ray is emitted directly from the
final excited state $f$ to a lower state $g$, which is observed experimentally
(see figure \ref{cascade1}(b)). The probability amplitude for this process is
\begin{equation}
a_{i\rightarrow f\rightarrow g}=\sum_{M_{f}}a_{i\rightarrow f}\;\left\langle
I_{g}M_{g}\mathbf{k}\sigma\left|  H_{\gamma}\right|  I_{f}M_{f}\right\rangle ,
\label{gad1}%
\end{equation}
where $\left\langle I_{g}M_{g}\mathbf{k}\sigma\left|  H_{\gamma}\right|
I_{f}M_{f}\right\rangle $ is the matrix element for the transition
$f\rightarrow g$ due to the emission of a photon with momentum \textbf{k} and
polarization $\sigma$. The operator $H_{\gamma}$ accounts for this transition.
The angular dependence of the $\gamma$-rays is given explicitly by the
spherical coordinates $\theta$ and $\phi$ of the vector \textbf{k}.

Since the angular emission probability will be normalized to unity, we can
drop constant factors and write it as (an average over initial spins is
included)
\begin{equation}
W\left(  \theta\right)  =\sum_{M_{i},\;M_{g},\sigma}\left|  a_{i\rightarrow
g}\right|  ^{2}=\sum_{M_{i},\;M_{g},\sigma}\left|  \sum_{M_{f}}a_{i\rightarrow
f}\;\left\langle I_{g}M_{g}\mathbf{k}\sigma\left|  H_{\gamma}\right|
I_{f}M_{f}\right\rangle \right|  ^{2}. \label{gad2}%
\end{equation}

The transition operator $H_{\gamma}$\ can be written as
\begin{equation}
H_{\gamma}=\sum_{lm^{{}}}H_{\gamma}^{\left(  lm\right)  }=\sum_{lm}%
\hat{\mathcal{O}}_{lm}^{(nuc)}\otimes\hat{\mathcal{O}}_{lm}^{\left(
\gamma\right)  }, \label{gad2_b}%
\end{equation}
where the first operator in the sum acts between nuclear states, whereas the
second operator acts between photon states of well defined angular momentum,
$l,m$.

Expanding the photon state $\left|  \mathbf{k}\sigma\right\rangle $ in a
complete set $\left|  lm\right\rangle $\ of the photon angular momentum, and
using the Wigner-Eckart theorem (angular momentum notation of ref.
\cite{Ed60}), one gets
\begin{align}
\left\langle I_{g}M_{g}\mathbf{k}\sigma\left|  H_{\gamma}\right|  I_{f}%
M_{f}\right\rangle  &  =\sum_{l,m}\left\langle \mathbf{k}\sigma
|lm\right\rangle \left\langle I_{g}M_{g}\left|  H_{\gamma}^{\left(  lm\right)
}\right|  I_{f}M_{f}\right\rangle \nonumber\\
&  =\left(  -1\right)  ^{I_{f}-M_{f}}\sum_{l,m}\left(
\begin{array}
[c]{ccc}%
I_{f} & l & I_{g}\\
-M_{f} & m & M_{g}%
\end{array}
\right)  \left\langle \mathbf{k}\sigma|lm\right\rangle \left\langle
I_{g}\left\|  H_{\gamma}^{\left(  l\right)  }\right\|  I_{f}\right\rangle .
\label{gad3}%
\end{align}

One can rewrite $\left|  \mathbf{k}\sigma\right\rangle $ in terms of $\left|
\mathbf{z}\sigma\right\rangle $, i.e., in terms of a photon propagating in the
z-direction. This is accomplished by rotating $\left|  \mathbf{k}
\sigma\right\rangle $ to the z-axis, using of the rotation matrix \cite{BS94},
$D_{mm^{\prime}}^{l}$, i.e.,
\begin{equation}
\left\langle \mathbf{k}\sigma|lm\right\rangle =\sum_{m^{\prime}}D_{mm^{\prime
}}^{l}\left(  \mathbf{z\longrightarrow k}\right)  \left\langle \mathbf{z}
\sigma|lm^{\prime}\right\rangle . \label{gad4}%
\end{equation}

Expanding the photon field in terms of angular momentum eigenfunctions, one
can show that \cite{FS65,EG88}
\begin{equation}
\left\langle \mathbf{z}\sigma|lm\pi\right\rangle =\left\{
\begin{array}
[c]{c}%
\sqrt{\frac{2l+1}{2}}\delta_{\sigma m}\;\;\;\;\;\mathrm{for}\;\;\;\pi=E\\
\sqrt{\frac{2l+1}{2}}\sigma\delta_{\sigma m}\;\;\;\;\;\mathrm{for}\;\;\;\pi=M.
\end{array}
\right.  \label{gad5}%
\end{equation}

One has now to express the operator $\hat{\mathcal{O}}_{l^{\prime}m^{\prime}%
}^{\left(  \gamma\right)  }$ in eq. \ref{gad2_b} in terms of the electric and
magnetic multipole parts of the photon field. This problem is tedious but
straightforward \cite{BR53}). Inserting eqs. \ref{gad4} and \ref{gad5} in eq.
\ref{gad3}, yields (neglecting constant factors)
\begin{align}
\left\langle I_{g}M_{g}\mathbf{k}\sigma\left|  H_{\gamma}\right|  I_{f}%
M_{f}\right\rangle  &  =\sum_{l,m}\left(  -1\right)  ^{I_{g}-l+M_{f}}%
\sqrt{\left(  2l+1\right)  \left(  2I_{f}+1\right)  }\left(
\begin{array}
[c]{ccc}%
I_{g} & l & I_{f}\\
M_{f} & m & -M_{f}%
\end{array}
\right) \nonumber\\
&  \times D_{m\sigma}^{l}\left(  \mathbf{z\longrightarrow k}\right)  \left[
\Delta_{El}+\sigma\Delta_{Ml}\right]  , \label{gad6}%
\end{align}
where $\Delta_{\pi l}$ is given by eq. \ref{gad0_1}.

Inserting eq. \ref{gad6} into eq. \ref{gad2} one gets a series of sums over
the intermediate values of the angular momenta
\begin{align}
W\left(  \theta\right)   &  = \sum_{ \overset{M_{i},M_{g},\sigma,M_{f},}%
{M_{f}^{\prime}, l,m,l^{\prime},m^{\prime}} } a_{i\longrightarrow
f}a_{i\longrightarrow f^{\prime}}^{\ast}\sqrt{\left(  2l+1\right)  \left(
2l^{\prime}+1\right)  \left(  2I_{f}+1\right)  }\left(
\begin{array}
[c]{ccc}%
I_{g} & l & I_{f}\\
M_{g} & m & -M_{f}%
\end{array}
\right) \nonumber\\
&  \times\left(  -1\right)  ^{M_{f}+M_{f}^{\prime}-l-l^{\prime}}\left(
\begin{array}
[c]{ccc}%
I_{g} & l^{\prime} & I_{f}\\
M_{g} & m^{\prime} & -M_{f}^{\prime}%
\end{array}
\right)  D_{m\sigma}^{l}\left[  D_{m^{\prime}\sigma}^{l^{\prime}}\right]
^{\ast}\Delta_{l}\Delta^{*}_{l^{\prime}}\;, \label{gad7_b}%
\end{align}
where $\Delta_{l}=\Delta_{El}+l\Delta_{Ml}$. The product $\Delta_{l}\Delta
^{*}_{l^{\prime}}$ is always real since $(-1)^{s(l)}=\Pi$ (the parity).

Assuming that the particles are detected symmetrically around the z-axis one
can integrate over $\phi_{particle}$, what is equivalent to integrating, or
averaging, over $\phi_{\gamma}$. This yields the following integral
\begin{align}
&  \int d\phi D_{m\sigma}^{l}\left[  D_{m^{\prime}\sigma}^{l^{\prime}}\right]
^{\ast}=\nonumber\\
&  \delta_{mm^{\prime}}\left(  -1\right)  ^{m-\sigma}\sum_{j}\frac{2j+1}%
{\sqrt{4\pi}}\left(
\begin{array}
[c]{ccc}%
l & j & l^{\prime}\\
m & 0 & -m
\end{array}
\right)  \left(
\begin{array}
[c]{ccc}%
l & j & l^{\prime}\\
\sigma & 0 & -\sigma
\end{array}
\right)  P_{j}\left(  \cos\theta\right)  . \label{gad7_c}%
\end{align}
To simplify further eq. \ref{gad7_b} we use (see ref. \cite{AW65}, p. 441, eq.
II.A.61)
\begin{align}
&  \sum_{M_{g}}\left(
\begin{array}
[c]{ccc}%
I_{g} & l & I_{f}\\
M_{g} & m & -M_{f}%
\end{array}
\right)  \left(
\begin{array}
[c]{ccc}%
I_{g} & l^{\prime} & I_{f}^{\prime}\\
M_{g} & m^{\prime} & -M_{f}^{\prime}%
\end{array}
\right) \nonumber\\
&  =\left(  -1\right)  ^{2l^{\prime}-I_{g}}\sum_{k,\kappa}\left(  -1\right)
^{k+m-M_{f}^{\prime}}\left(  2k+1\right)  \left(
\begin{array}
[c]{ccc}%
l & l^{\prime} & k\\
m & -m^{\prime} & \kappa
\end{array}
\right)  \left(
\begin{array}
[c]{ccc}%
I_{f} & I_{f}^{\prime} & k\\
M_{f} & -M_{f}^{\prime} & \kappa
\end{array}
\right)  \left\{
\begin{array}
[c]{ccc}%
l & l^{\prime} & k\\
I_{f}^{\prime} & I_{f} & I_{g}%
\end{array}
\right\}  \label{gad7_d}%
\end{align}
and
\[
\sum_{\sigma=\left(  -1,1\right)  }\left(
\begin{array}
[c]{ccc}%
l & j & l^{\prime}\\
\sigma & 0 & -\sigma
\end{array}
\right)  \Delta_{\pi l}\Delta^{*}_{\pi l^{\prime}}=\left\{
\begin{array}
[c]{c}%
2\left(
\begin{array}
[c]{ccc}%
l & j & l^{\prime}\\
1 & 0 & -1
\end{array}
\right)  \Delta_{\pi l}\Delta^{*}_{\pi l^{\prime}}\;,\;\;\;\;\;\mathrm{for}%
\;\;\;j=\mathrm{even}\\
0\;,\;\;\;\;\;\;\;\;\;\mathrm{for}\;\;\;\;\;j=\mathrm{odd},
\end{array}
\right.
\]
where use has been made of the parity selection rule
\[
\Pi_{1}\Pi_{2}=\left\{
\begin{array}
[c]{c}%
\left(  -1\right)  ^{l}\;,\;\;\;\mathrm{for}%
\;\;\;\mathrm{electric\;transitions}\\
\left(  -1\right)  ^{l+1}\;,\;\;\;\mathrm{for}%
\;\;\;\mathrm{magnetic\;transitions \ .}%
\end{array}
\right.
\]

Eq. \ref{gad7_b} becomes
\begin{align}
W\left(  \theta\right)   &  =\sum_{ \overset{M_{i},k,\kappa, M_{f}%
,M_{f}^{\prime},}{l,l^{\prime},m,m^{\prime}} } \left(  -1\right)
^{2m^{\prime}+k+M_{f}}a_{i\longrightarrow f}a_{i\longrightarrow f^{\prime}%
}^{\ast}\sqrt{\left(  2l+1\right)  \left(  2l^{\prime}+1\right)  \left(
2I_{f}+1\right)  }\nonumber\\
&  \times\left(  2j+1\right)  \left(  2k+1\right)  \left(
\begin{array}
[c]{ccc}%
I_{f} & I_{f} & k\\
M_{f} & -M_{f}^{\prime} & \kappa
\end{array}
\right)  \left(
\begin{array}
[c]{ccc}%
I_{f} & l^{\prime} & k\\
M_{f} & -m^{\prime} & \kappa
\end{array}
\right) \nonumber\\
&  \times\left(
\begin{array}
[c]{ccc}%
l & j & l^{\prime}\\
1 & 0 & -1
\end{array}
\right)  \left(
\begin{array}
[c]{ccc}%
l & j & l^{\prime}\\
m & 0 & -m^{\prime}%
\end{array}
\right)  \left\{
\begin{array}
[c]{ccc}%
l & l^{\prime} & k\\
I_{f}^{\prime} & I_{f} & I_{g}%
\end{array}
\right\}  \Delta_{l}\Delta^{*}_{l^{\prime}}P_{j}\left(  \cos\theta\right)  \;.
\end{align}

Using
\[
\sum_{m,m^{\prime}}\left(  -1\right)  ^{2m^{\prime}}\left(
\begin{array}
[c]{ccc}%
l & j & l^{\prime}\\
m & 0 & -m^{\prime}%
\end{array}
\right)  \left(
\begin{array}
[c]{ccc}%
l & j & l^{\prime}\\
1 & 0 & -1
\end{array}
\right)  =\left(  -1\right)  ^{l+l^{\prime}+k}\frac{1}{2k+1}\delta_{kj}%
\delta_{\kappa0},
\]
one gets
\begin{align*}
W\left(  \theta\right)   &  =\sum_{ \overset{k=\mathrm{even},M_{i},}{M_{f},
l,l^{\prime}} } \left(  -1\right)  ^{l+l^{\prime}+k}\sqrt{\left(  2l+1\right)
\left(  2l^{\prime}+1\right)  \left(  2I_{f}+1\right)  }\left(  2k+1\right)
\left|  a_{i\longrightarrow f}\right|  ^{2}\\
&  \times\left(
\begin{array}
[c]{ccc}%
I_{f} & I_{f} & k\\
M_{f} & -M_{f} & 0
\end{array}
\right)  \left(
\begin{array}
[c]{ccc}%
l & j & l^{\prime}\\
1 & 0 & -1
\end{array}
\right)  \left\{
\begin{array}
[c]{ccc}%
l & l^{\prime} & k\\
I_{f}^{\prime} & I_{f} & I_{g}%
\end{array}
\right\}  \Delta_{l}\Delta^{*}_{l^{\prime}}P_{j}\left(  \cos\theta\right)  ,
\end{align*}
or in a more compact form
\begin{equation}
W\left(  \theta\right)  =\sum_{ \overset{ k=\mathrm{even},M_{i},M_{f},}{
l,l^{\prime}} } \left(  -1\right)  ^{M_{f}}\left|  a_{i\longrightarrow
f}\right|  ^{2} F_{k}\left(  l,l^{\prime},I_{g},I_{f}\right)  \left(
\begin{array}
[c]{ccc}%
I_{f} & I_{f} & k\\
M_{f} & -M_{f} & 0
\end{array}
\right)  \sqrt{2k+1}P_{k}\left(  \cos\theta\right)  \Delta_{l}\Delta^{*}
_{l^{\prime}}\;,
\end{equation}
where
\begin{align}
F_{k}\left(  l,l^{\prime},I_{g},I_{f}\right)   &  =\left(  -1\right)
^{I_{f}-I_{g}-1}\sqrt{\left(  2l+1\right)  \left(  2l^{\prime}+1\right)
\left(  2I_{f}+1\right)  \left(  2k+1\right)  }\nonumber\\
&  \times\left(
\begin{array}
[c]{ccc}%
l & l^{\prime} & k\\
1 & -1 & 0
\end{array}
\right)  \left\{
\begin{array}
[c]{ccc}%
l & l^{\prime} & k\\
I_{f} & I_{f} & I_{g}%
\end{array}
\right\}  \ .
\end{align}

The angular distribution of $\gamma$-rays described above is in the reference
frame of the excited nucleus. To obtain the distribution in the laboratory one
has to perform the transformation
\begin{equation}
\theta_{L}=\arctan\left\{  {\frac{\sin\theta}{\gamma\left[  \cos\theta+
\beta\right]  }}\right\}  \ ,
\end{equation}
and
\begin{equation}
W (\theta_{L}) = \gamma^{2} \left(  1+ \beta\cos\theta\right)  ^{2} W
(\theta)\ ,
\end{equation}
where $\gamma$ is given by eq. \ref{transf3}, and $\beta=\sqrt{1-1/\gamma^{2}%
}$. The photon energy in the laboratory is $E^{ph}_{L}=\gamma E^{ph}_{cm}
\left(  1 +\beta\cos\theta\right)  $.

\subsection{Computer program and user's manual}

All nuclear quantities, either known from experiments or calculated from a
model, as well as the conditions realized in the experiment, are explicitly
specified as input parameters. The program DWEIKO then computes the optical
potentials (if required), differential cross section for elastic scattering,
and Coulomb + nuclear excitation probabilities and cross sections, as well as
the angular distribution of the $\gamma$-rays.

The units used in the program are fm (femtometer) for distances
and MeV for energies. The output cross sections are given in
millibarns.

\subsubsection{Input parameters}

{\small To avoid exceeding use of computer's memory, the file DWEIKO.DIM
contains the dimension of the arrays and sets in the maximum number of levels
(NMAX), maximum total number of magnetic substates, (NSTMAX), maximum number
of impact parameters (NBMAX), and maximum number of coordinates points used in
the optical potentials and absorption factors, (NGRID). A good estimate is
NSTMAX = $(2J_{max}+1)$NST, where $J_{max}$ is the maximum angular momentum of
the input states}.

{\small Most integrals are performed by the 1/3-Simpson's integration rule. It
is required that NGRID be a \textit{even} number, since an extra point
(origin) is generated in the program. }

{\small The input file allows for comment lines. These should start with a
`$\#$' sign. }

{\small The file DWEIKO.IN contains all other input parameters. These are }

\begin{enumerate}
\item {\small AP, ZP, AT, ZT, which are the projectile and the target mass and
charge numbers, respectively. }

\item {\small ECA, the bombarding energy per nucleon in MeV. }

\item {\small EX(j) and SPIN(j): the energy and spins of the individual states
j. }

\item {\small MATE1(j,k), MATE2(j,k), MATE3 (j,k), MATM1(j,k), MATM2(j,k) the
reduced matrix elements for E1, E2, E3 and M1, M2 excitations, $j\rightarrow
k,$ \ (as defined in (\ref{Mfi1})), in $e$ fm (E1,M1), $e$ fm$^{2}$ (E2,M2),
and $e$ fm$^{4}$ (E3) units. }

\item {\small F(0,j), F(1,j), F(2,j), F(3,j), the fractions of sum rule of the
deformation parameters for monopole, dipole, quadrupole, and octupole nuclear
excitations, entering eq. (\ref{VfiN2}). }

{\small To simplify the input, the deformation parameters are calculated
internally in DWEIKO using the sum rules \ref{deform1}-\ref{deform11}, with
$E_{x}$ replaced by the energy of the corresponding state, EX(j). The user
needs to enter the fraction of those sum rules exhausted by the state j, i.e.,
the program uses $\delta_{\lambda}^{\prime}= f_{\lambda}\delta_{\lambda}$,
with ($0\leq f_{\lambda}\leq1$) entered by the user, and $\delta_{\lambda}$
given by eqs. \ref{deform1}-\ref{deform11}. }
\end{enumerate}

{\small The input cards in file DWEIKO.IN are organized as following: }

\begin{enumerate}
\item {\small AP, ZP, AT, ZT, ECA}

{\small Charges and masses (AP,ZP,AT,ZT), bombarding energy per nucleon in
MeV/nucleon. }

\item {\small IW, IOPM, IOELAS, IOINEL, IOGAM }

{\small IW=0(1) for projectile (target) excitation. }

{\small IOPM=1(0) for output (none) of optical model potentials. }

{\small IOELAS=(0)[1]{2} for (no output) [center of mass] {laboratory} elastic
scattering cross section. }

{\small IOINEL=(0)[1]{2} for (no output) [center of mass] {laboratory}
inelastic scattering cross section. }

{\small IOGAM=(0)[1]{2} for (no output) [output of statistical tensors]
{output of gamma-ray angular distributions}. }

{\small The statistical tensors are calculated for each impact parameter, so
that one can use them in the computation of $dP_{\gamma N\rightarrow
M}(b)/d\Omega_{\gamma}=P_{N}(b).dW_{\gamma N\rightarrow M}/d\Omega_{\gamma}$.
If IW = 0, a transformation to the laboratory system is performed. }

\item {\small NB, ACCUR, BMIN, IOB }

{\small NB = number of impact parameter points (NB $\leq$ NBMAX). }

{\small ACCUR = accuracy required for the time integration of the CC-equations
for each impact parameter. A reasonable value is ACCUR = 0.001, i.e., 0.1\%. }

{\small BMIN=minimum impact parameter (enter 0 for default. The program will
integrate from $b=0$, with strong absorption included). }

{\small IOB=1(0) prints (does not print) out impact parameter probabilities. }

\item {\small IOPW, IOPNUC }

{\small IOPW is a switch for the optical potential model (OPM). }

{\small IOPW=0 (no OMP, IOELAS=0), 1 (Woods-Saxon), 2 (read), 3 (t-$\rho\rho$
folding potential), 4 (M3Y folding potential). }

{\small IOPNUC=0 (no nuclear), 1 (vibrational excitations). }

{\small If the optical potential is provided (IOPW=2), it should be stored in
`optw.in' in rows of R x Real[U(R)] x Imag[U(R)]. The program makes an
interpolation to obtain intermediate values. }

{\small The first line in `optw.in' is the number of rows (maximum=NGRID). }

\item {\small V0 [MeV], r0 [fm], d [fm], VI [MeV], r0$\_$I [fm], dI [fm] }

{\small If IOPW=1, enter V0 [VI] = real [imaginary] part ($>$0) of Woods-Saxon
potential, r0 [r0$\_$I] = radius parameter (R = r0 * (AP$^{1/3}$ + AT$^{1/3}%
$), d [dI] = diffuseness, }

{\small If IOPW is not equal to 1, place a `$\#$' sign at the beginning of
this line, or delete it. }

\item {\small VS0 [MeV], r0$\_$S [fm], dS [fm], V$\_$surf [fm], a$\_$surf [fm]
}

{\small If IOPW=1 and AP, or AT, equal to one (proton), enter here spin-orbit
part. If not, place a '$\#$' sign at the beginning of this line, or delete it.
}

{\small VS0 = depth parameter of the spin-orbit potential ($>$0), r0$\_$S =
radius parameter, dS = diffuseness, V$\_$surf = depth parameter of the surface
potential ($>$0), a$\_$surf = diffuseness, }

\item {\small Wrat }

{\small If IOPW=4, enter Wrat = ratio of imaginary to real part of M3Y
interaction. }

{\small If IOPW is not equal to 4, place a `$\#$' sign at the beginning of
this line, or delete it. }

\item {\small THMAX, NTHETA }

{\small If IOELAS=1,2 or IOINEL=1,2 enter here THMAX, maximum angle (in
degrees and in the center of mass), and NTHETA ($\leq$ NGRID), the number of
points in the scattering angle. }

{\small If IOELAST or IOINEL are not 1, or 2, place a `$\#$' sign at the
beginning of this line, or delete it. }

\item {\small JINEL }

{\small If IOINEL=1 enter the state (JINEL) for the inelastic angular
distribution ($1 < $ JINEL $\leq$ NST) }.

{\small If IOINEL is not 1, or 2 place a `$\#$' sign at the beginning of this
line, or delete it. }

\item {\small NST }

{\small NST ($\leq$ NMAX) is the number of nuclear levels. }

\item {\small I, EX(I), SPIN(I) }

{\small Input of state labels (I), energy EX(I), and angular momentum SPIN(I).
I ranges from 1 to NST and should be listed in increasing value of energies. }

\item {\small I, J, E1[e fm], E2[e fm$^{2}$], E3[e fm$^{3}$], M1[e fm], M2[e
fm$^{2}$] }

{\small Reduced matrix elements for E1, E2, E3, M1 and M2 excitations:
$<I_{j}||O(E/M;L)||I_{i}>$, j $\geq$ i , for the electromagnetic transitions.
Matrices for reorientation effects, i$\rightarrow$i, can also be given. }

{\small Add a row of zeros at the end of this list. If no electromagnetic
excitation is wanted just enter a row of zeros.}

\item {\small J, F0, F1, F2, F3 }

{\small If IOPNUC=1 enter sum rule fraction of nuclear deformation parameters
for monopole, dipole, quadrupole nuclear excitations
(ALPHA0,DELTE1,DELTE2,DELTE3) for each excited state J. }

{\small If IOPNUC=0 insert a comment card (`$\#$') in front of each entry row,
or delete them. }

\item {\small IFF, IGG, THMIN, THMAX, NTHETA }

{\small If IOGAM=2, enter IFF,IGG = initial and final states (IFF $>$ IGG) for
the gamma transition. }

{\small THMIN, THMAX = minimum and maximum values of gamma-ray angles (in
degrees) in the laboratory frame. }

{\small NTHETA = number of angle points ($\leq$ NGRID). }
\end{enumerate}

\subsubsection{{\protect\normalsize Computer program}}

{\normalsize The program starts with a catalogue of the nuclear levels by
doing a correspondence of integers to each magnetic substate. J = 1
corresponds to the lowest energy level, with the magnetic quantum number
$M_{1}=-I_{1}$. J increases with $M_{1}$ and so on for the subsequent levels.
}

{\normalsize A mesh in impact parameter is done, reserving half of the impact
parameter points, i.e., $NB/2$, to a finer mesh around the grazing impact
parameter, defined as $b_{0}=1.2\left(  A_{P}^{1/3}+A_{T}^{1/3}\right)
\;$fm$.$ The interval $b_{0}/3\;fm\leq b\leq2b_{0}\;fm$ is covered by this
mesh. A second mesh, with the other half of points, extends from
$b=2b_{0}\;fm$ to $b=200\;fm.$ Except for the very low excitation energies
($E_{j}\ll1$ MeV), combined with very large bombarding energies ($\gamma\gg
1$), this upper value of $b$ corresponds to very small excitation
probabilities, and the calculation can be safely stopped. The reason for a
finer mesh at small impact parameters is to get a good accuracy at the region
where both nuclear, Coulomb, and absorption factors play equally important
roles. At large impact parameters the probabilities fall off smoothly with
$b$, justifying a wider integration step. }

{\normalsize A mesh of NGRID points in polar coordinates is implemented to
calculate the nuclear excitation potentials and absorption factors, according
to the equations presented in sections 2.2 and 2.3. The first and second
derivatives of the optical potentials are calculated by the routine
DERIVATIVE. A 6-point formula is used for the purpose. The routines TWOFOLD
computes the folding over the densities, as used in eq. (\ref{abs2}). Routines
RHOPP and RHONP generate the liquid drop densities, and the routine PHNUC
computes the eikonal integral appearing in eq. (\ref{abs}). }

{\normalsize Repeated factors for the nuclear and for the Coulomb potentials
are calculated in the main program and stored in the main program with the
arrays PSITOT and PSINUC. These arrays are carried over in a common block to
the routine VINT which computes the function $Q_{N\lambda\mu}^{(kj)}(\xi_{kj},
\tau)+Q_{C\pi\lambda\mu}(\xi_{kj},\tau)$, used in eq. (\ref{ats2}). }

{\normalsize The optical potentials are read in routine OMP$\_$READ, or
calculated in routines OMP$\_$WS (Woods-Saxon), OMP$\_$DEN (t-$\rho\rho$) and
OMP$\_$M3Y (M3Y). Routine TM3Y sets the M3Y interaction. DEFORM calculates the
effective potentials in the Bohr-Mottelson model. SIGNNE and PHNNE return the
nucleon-nucleon cross sections and the parameters of the nucleon-nucleon
scattering amplitude \ref{opt6}. PHNUCF calculates the eikonal phase in the
``t-$\rho\rho$" approximation with the help of the Fourier transform of the
ground state densities, provided by FOURIER. }

{\normalsize The time integrals are performed by means of an adaptive
Runge-Kutta method. All routines used for this purpose have been taken from
the Numerical Recipes \cite{Numrep}. They are composed by the routines ODEINT,
RKQS, RKCK, and RK4. The routine ODEINT varies the time step sizes to achieve
the desired accuracy, controlled by the input parameter ACCUR. The right side
of (\ref{ats2}) is computed in the routine DCADT, used externally by the
fourth-order Runge-Kutta routine RK4. RKCK is a driver to increase time steps
in RK4, and RKQS is used in ODEINT for the variation of step size and accuracy
control. The main program returns a warning if the summed errors for all
magnetic substates is larger than 10 $\times$ ACCUR. }

{\normalsize Elastic scattering is calculated within the routine
ELAST for nucleus-nucleus, and ELASTP for proton-nucleus,
collisions. Inelastic scattering is calculated within the routine
INELAST, and $\gamma$-ray angular distributions are calculated in
the routine GAMDIS. }

{\normalsize The routine THREEJ computes Wigner-3J coefficients
(and Clebsh-Gordan coefficients), RACAH the 6j-symbols, or Racah
coefficients, YLM is used to compute the spherical harmonics,
LEGENGC the Legendre polynomials, and BESSJ0 (BESSJ1) [BESSJN] the
Bessel function $J_{0}$ ($J_{1}$) [$J_{n}$].}

{\normalsize The routines SPLINE and SPLINT perform a spline interpolation of
the excitation amplitudes, before they are used for integration by means of
the routines QTRAP and QSIMP, from Numerical Recipes \cite{Numrep}. }

\section{{\protect\normalsize Test input and things to do}}

{\normalsize A test input is shown below. It corresponds to the excitation of
giant resonance states in Pb by means of the reaction $^{17}$O (84
MeV/nucleon) + $^{208}$Pb. Assuming that an isolated state is excited, and
that it exhausts fully the sum rules, one gets ($B(E\lambda)\equiv B(E\lambda,
E_{x})$)
\begin{equation}
B(E1)= {\frac{9 }{4\pi}} \ {\frac{\hbar^{2} }{2 m_{N}}} \ {\frac{NZ}{AE_{x}}}
\ e^{2} \ ,
\end{equation}
and, for ($\lambda> 1$)
\begin{equation}
B(E\lambda)= {\frac{3 }{4\pi}}\ \lambda(2\lambda+1)\ {\frac{R^{2}}{E_{x}}}
\ {\frac{\hbar^{2}}{2m_{N}}} \ e^{2} \times\left\{
\begin{array}
[c]{ll}%
Z^{2}/A, & \mbox{ for isoscalar excitations; }\\
NZ/A, & \mbox{for isovector excitations; }
\end{array}
\right.
\end{equation}
From these values, the reduced matrix elements can be calculated from the
definition in eq. \ref{Bvalue}. }

{\normalsize \textbf{Example 1} - The input list below calculates the
excitation cross sections for the isovector giant dipole (IVGDR) and isoscalar
giant quadrupole (ISGQR) resonances in Pb, for the above mentioned reaction.
The user should first run this sample case and verify that the output numbers
check those in ref. \cite{Ba88}. In particular, compare the results for
elastic scattering with the upper figure 2 of ref. \cite{Ba88}. Also compare
the inelastic excitation of the IVGDR with the upper figure 3 of ref.
\cite{Ba88}. Following this it might be instructive to change the input of
energies, spins, and excitation strengths for low lying states, giant
resonances, etc. }

{\small $\#$\ \ \ Input of program `DWEIKO' }

{\small $\#$\ \ \ Ap\ \ \ Zp\ \ \ At\ \ \ Zt\ \ \ Einc[MeV/u] }

{\small \ \ \ 17 \ \ \ 8\ \ \ 208\ \ \ 82\ \ \ 84.\ \ \ }

{\small $\#$%
\ \ \ IW=0(1)\ \ \ IOPM=0(1)\ \ \ IOELAS=0(1)[2]\ \ \ IOINEL=0(1)[2]\ \ \ IOGAM=0(1)[2]
}

{\small \ \ \ 1\ \ \ 1\ \ \ 1\ \ \ 1\ \ \ 2 }

{\small $\#$\ \ \ NB\ \ \ ACCUR\ \ \ BMIN[fm]\ \ \ IOB=1(0) }

{\small \ \ \ 200\ \ \ 0.001\ \ \ 0.\ \ \ 0 }

{\small $\#$\ \ \ IOPW\ \ \ IOPNUC }

{\small \ \ \ 1\ \ \ 1 }

{\small $\#$\ \ \ V0 [MeV]\ \ \ r0[fm]\ \ \ d[fm]\ \ \ VI [MeV]\ \ \ r0$\_$I
[fm]\ \ \ dI [fm] }

{\small \ \ \ 50.\ \ \ 1.067\ \ \ 0.8\ \ \ 58.\ \ \ 1.067\ \ \ 0.8 }

{\small $\#$\ \ \ VS0 [MeV]\ \ \ r0$\_$S [fm]\ \ \ dS
[fm]\ \ \ V$_{\mathrm{surf}}$ [fm] \ \ \ d$_{\mathrm{surf}}$ }

{\small $\#$\ \ \ 15.\ \ \ 1.02\ \ \ 0.6\ \ \ 50. \ \ \ 0.8 }

{\small $\#$\ \ \ Wrat }

{\small $\#$ \ \ \ 1. }

{\small $\#$\ \ \ THMAX\ \ \ NTHETA }

{\small \ \ \ 6.\ \ \ 150 }

{\small $\#$\ \ \ JINEL }

{\small \ \ \ 3 }

{\small $\#$ \ \ \ NST }

{\small \ \ \ 3 }

{\small $\#$ \ \ \ I \ \ \ Ex[MeV]\ \ \ SPIN }

{\small \ \ \ 1\ \ \ 0\ \ \ 0 }

{\small \ \ \ 2\ \ \ 10.9\ \ \ 2 }

{\small \ \ \ 3\ \ \ 13.5\ \ \ 1 }

{\small $\#$\ \ \ i -$>$ j\ \ \ E1[e fm]\ \ \ E2[e fm**2]\ \ \ E3[e
fm**3]\ \ \ M1[e fm]\ \ \ M2[e fm** 2] }

{\small \ \ \ 1\ \ \ 2\ \ \ 0\ \ \ 76.8\ \ \ 0\ \ \ 0\ \ \ 0 }

{\small \ \ \ 1\ \ \ 3\ \ \ 9.3\ \ \ 0\ \ \ \ 0\ \ \ 0\ \ \ 0\ \ \ }

{\small \ \ \ 0\ \ \ 0\ \ \ 0\ \ \ \ 0\ \ \ 0\ \ \ \ \ 0\ \ \ 0 }

{\small $\#$\ \ \ j\ \ \ F0\ \ \ F1\ \ \ F2\ \ \ F3\ \ \ (NOTE: fractions of
sum rules for deformation parameters) }

{\small \ \ \ 2\ \ \ 0\ \ \ 0\ \ \ 0.6\ \ \ 0 }

{\small \ \ \ 3\ \ \ 0\ \ \ 1.1\ \ \ 0\ \ \ 0 }

{\small $\#$\ \ \ IFF\ \ \ IGG\ \ \ THMIN\ \ \ THMAX\ \ \ NTHETA }

{\small \ \ \ 2\ \ \ 1\ \ \ 20.\ \ \ 70.\ \ \ 150 }

{\normalsize \bigskip}

{\normalsize \textbf{Example 2} - Table III gives the results of an experiment
on Coulomb excitation of S and Ar isotopes \cite{heiko}. Choose the minimum
impact parameter, BMIN, so that it reproduces the maximum scattering angle
($\theta_{\mathrm{max}}=4.1^{\circ}$) in the experiment (use the formula
$b=a_{0} \cot({\theta/2})$ and read the last paragraph of section II). Using
the $B(E2)$ values in the table reproduce the cross section values (careful
with the units!). }

\begin{center}
{\normalsize
\begin{tabular}
[c]{|l|l|l|l|l|l|}\hline\hline
Secondary beam & $^{38}$S & $^{40}$S & $^{42}$S & $^{44}$Ar & $^{46}%
$Ar\\\hline
$E_{\mathrm{lab}}$ [MeV/nucleon] & 39.2 & 39.5 & 40.6 & 33.5 & 35.2\\\hline
Energy of the first excited state [MeV] & 1.286(19) & 0.891(13 & 0.890(15) &
1.144(17) & 1.554(26)\\\hline
$\sigma(E2;0_{g.s.}^{+}\rightarrow2_{1}^{+};\ \theta_{\mathrm{lab}}%
\geq4.1^{\circ})$ [mb] & 59(7) & 94(9) & 128(19) & 81(9) & 53(10)\\\hline
$B(E2;0_{g.s.}^{+}\rightarrow2_{1}^{+})$ [$e^{2}$ fm$^{4}$] & 235(30) &
334(36) & 397(63) & 345(41) & 196(39)\\\hline
\end{tabular}
}

{\normalsize Table III. Experimental results on Coulomb excitation of S and Ar
projectiles impinging on a $^{197}$Au target \cite{heiko}. }

\end{center}

{\normalsize \bigskip}

{\normalsize \textbf{Example 3} - Compare the results of this code with those
from the ECIS code \cite{Ra72,Ra81,Ra87}. But notice that the ECIS code has
relativistic corrections in kinematic variables only. The relativistic
dynamics in the Coulomb and nuclear interaction are not accounted for. Thus,
one should expect disagreements for high energy collisions ($E_{lab} \ge100$
MeV/nucleon). }

\section{{\protect\normalsize Output}}

{\small The output of DWEIKO are in the files }

\begin{enumerate}
\item {\small DWEIKO.OUT: Probabilities and cross sections; }

\item {\small DWEIKO$\_$OMP.OUT: Optical model potential; }

\item {\small DWEIKO$\_$ELAS.OUT: Elastic scattering cross section; }

\item {\small DWEIKO$\_$INEL.OUT: Inelastic scattering cross section; }

\item {\small DWEIKO$\_$STAT.OUT: Statistical tensors; }

\item {\small DWEIKO$\_$GAM.OUT: Angular distributions of gamma-rays. }
\end{enumerate}

\section{{\protect\normalsize Acknowledgments}}

{\normalsize We would like to thank Ian Thompson for helping us to
correct bugs in the program and for useful discussions. This
material is based on work supported by the National Science
Foundation under Grants No. PHY-0110253, PHY-9875122, PHY-007091
and PHY-0070818. }

{
}

\end{document}